# A pedagogic review on designing model topological insulators

Tanmoy Das

*Department of Physics, Indian Institute of Science, Bangalore- 560012, India.*

Following the centuries old concept of the quantization of flux through a Gaussian curvature (Euler characteristic) and its successive dispersal into various condensed matter properties such as quantum Hall effect, and topological invariants, we can establish a simple and fairly universal understanding of various modern topological insulators (TIs). Formation of a periodic lattice (which is a non-trivial Gaussian curvature) of 'cyclotron orbits' with applied magnetic field, or 'chiral orbits' with fictitious 'momentum space magnetic field' (Berry curvature) guarantees its flux quantization, and thus integer quantum Hall (IQH), and quantum spin-Hall (QSH) insulators, respectively, occur. The bulk-boundary correspondence associated with all classes of TIs dictates that some sort of pumping or polarization of a 'quantity' at the *boundary* must be associated with the flux quantization or topological invariant in the *bulk*. Unlike charge or spin polarizations at the edge for IQH and QSH states, the time-reversal (TR) invariant $Z_2$ TIs pump a mathematical quantity called 'TR polarization' to the surface. This requires that the valence electron's wavefunction (say, $\psi_\uparrow(\mathbf{k})$) switches to its TR conjugate ($\psi_\downarrow^*(-\mathbf{k})$) odd number of times in half of the Brillouin zone. These two universal features can be considered as 'targets' to design and predict various TIs. For example, we demonstrate that when two adjacent atomic chains or layers are assembled with opposite spin-orbit coupling (SOC), serving as the TR partner to each other, the system naturally becomes a $Z_2$ TI. This review delivers a holistic overview on various concepts, computational schemes, and engineering principles of TIs.

## CONTENTS

Email: tnmydas@gmail.com

## I. Introduction

Phase transition is distinguished by a change in symmetry, involving either a reduction or addition of symmetry in the ground state. A reduction of symmetry, which commonly involves translational, time-reversal (TR), rotational, gauge symmetries, among others, leads to a classical/quantum phase transition, and is defined by an order parameter within the Landau's paradigm. On the contrary a topological phase is defined by the emergence of a new quantum number (such as Chern number, $Z_2$ invariant), arising from the geometry or topology of the band structure. The topological invariant can be understood from a pure mathematical formalism of the Euler characteristic or Euler number. This implies that the net flux through a Gaussian curvature is always quantized. This is precisely what happens, according to Laughlin's argument,[1] in two-dimensional (2D) lattices (which can be represented by a torus - a Gaussian curvature) when magnetic field is applied perpendicular to it. In this case, the magnetic flux through the 2D system or torus remains quantized, giving rise to integer quantum Hall (IQH) effect. This is the first realization of topological invariant in condensed matter science.

IQH is a well understood phenomenon, with different ways to quantify its topological invariants. For example, Thouless, Kohmoto, Nightingale, and Nijs (TKNN) have shown that the IQH effect can also be understood from the Berry phase paradigm in the momentum space.[2] The corresponding topological invariant is thus known as TKNN number or Chern number. Haldane proposed a pioneering idea to obtain quantum Hall (QH) effect without external magnetic field in a honeycomb lattice.[3] Honeycomb lattice has two different sublattices. With the application of an external gauge field, the intra-sublattice hoppings possess chiral motion, and the chirality of two sublattices becomes opposite to each other. Different intra-sublattice electron hopping commence counter-propagating triangular 'cyclotron orbits', each threading opposite 'magnetic fields'. As we will go along in this review, we will identify that the formation of localized 'cyclotron orbitals' in 2D lattice without a magnetic field, which we call "chiral orbits", is a key ingredient to obtain QH effect. Each such "chiral orbit" now encloses integer flux quanta, in the same language of the IQH effect, and due to TR symmetry breaking a net QH effect survives.

The next development to the TI field was put forward by Kane and Mele in 2005 for obtaining TR invariant TIs, as known by $Z_2$ TI.[4] They realized that Haldane's 'gauge field' can be achieved by spin-orbit coupling (SOC). Owing to the spin-momentum locking, the right- and left-moving electrons have opposite spin polarizations. Since the TR symmetry is intact here, the flux passing through different spin-resolved 'chiral orbits' in a 2D lattice are exactly equal but opposite. This makes the net flux to be zero, leading to no charge pumping to the edge, but the difference between the two fluxes is finite, giving rise to a net spin-resolved QH effect, as known by quantum spin-Hall (QSH) effect. This is the foundation of the 2D $Z_2$ TI.

The generalization of the $Z_2$ topological invariant to 3D cannot, however, be easily done in terms of 'chiral orbit' formations, except in special cases of layered systems and heterostructures.[5] There are several mathematical formulations of the $Z_2$ invariant[4,6–14]. Among which, the Kane-Mele method of TR polarization is widely used.[4] They proposed a derivation of the bulk $Z_2$ topological invariant from the bulk-boundary correspondence which is a necessary condition for all topological invariants. Recall that in the case of IQH and QSH insulators, charge and spin are accumulated at the edge. Based on this, they enquired that something similar must be pumped to the boundary in $Z_2$ TIs. Since QSH insulator also belongs to the $Z_2$ class, the quantity that is pumped to the edge must accommodate spin as a subset. Since the spin-up state at the +**k** and spin-down state at –**k** are TR conjugates to each other, Kane-Mele proposed that a more general mathematical quantity, called 'TR polarization', is accumulated at the edge in this case. Here electrons possessing a particular wavefunction and their TR conjugate partners move to different edges. This incipiently requires that the electron exchanges it TR partner odd number of times in traversing half-of the Brillouin zone (BZ). In what follows, $Z_2$ topological invariant is nothing but counts of the number TR partner exchanges; odd number corresponds to $Z_2$ invariant $\nu = 1$, while even number implies $\nu = 0$. Since the TR symmetric topological invariant only takes two values, it has $Z_2$ symmetry. But TR breaking IQH state can take arbitrarily large Chern number.

Chirality of electrons is an essential ingredient for TIs, which is obtained either by magnetic field, or in bipartite lattice (such as staggered hopping in Su-Schrieffer-Heeger (SSH) lattice,[15] or graphene) or via SOC. Such chirality can also arise from the orbital texture inversion between even and odd parity orbitals at the TR invariant points, leading to a distinct class of spinless topological and Dirac materials.[16,17] In simple term, chirality means that the electron's hopping in lattice must be complex. As the electron's hopping encloses a 'chiral orbit', the complex phase manifests into a magnetic field at the center – either applied, or self-generated (Berry curvature). For $Z_2$ class, the TR partner switching, discussed in the previous paragraph, is nothing but the exchange of electron's phase to its complex conjugate odd number of times in half-of the BZ.

Subsequently, Fu-Kane simplified the calculation of $Z_2$ invariant by using the parity analysis.[7] They showed that if a system has both TR and inversion symmetries, the $Z_2$ topological invariant can be computed simply by counting how many times the electron exchanges it parity at the TR invariant momenta. If the valence band is not fully defined by a single parity, rather it exchanges parities odd number of times with the conduction band (as in the case of TR partner exchange), it gives a non-trivial topological invariant. This is also equivalent to chirality inversion in special cases as we will demonstrate in our engineering procedures. The band gap between the opposite parity conduction and valence bands at the TR invariant momenta also serves as the 'negative' Dirac mass in the Dirac equation.[9,18] The consequence of this band inversion is very rich, allowing protected gapless surface or edge states with Dirac cone. This provides an alternative springboard to obtain numerous exotic properties originally proposed by solving Dirac equation in the high-energy theories.

This simple theory leads to the search of materials with band inversions at the TR invariant k-points as a simple tool to identify $Z_2$ TI. The mechanism of parity inversion is not unique, and can in-principle vary to a wide range of tuning parameters as well as electron-electron interactions. Among them, SOC is responsible for the band inversion in most of the known TIs. Initially, the search for TI had been very much 'blind-folded', seeking materials with odd number of band inversions triggered by SOC.[6,9,18,19] Subsequently, more advanced methods of TI materials genome such as 'adiabatic transformation' method was developed,[20] which can be applied to systems without inversion symmetry. In this method, one starts with a known TI, and continuously tunes the atomic number of the constituent elements, and arrives at a new material. In this process if the band gap does not close and reopen at the TR invariant points, the new material must also be a non-trivial TI. These two methods have enabled the discoveries of a rich variety of TI materials.

Subsequently, various distinct classes of TI are predicted and discovered. For example, mirror symmetry and *p*-wave superconducting pairing symmetry can lead to two distinct classes of TIs, called topological crystalline insulator,[21] and topological superconductor,[22,23] respectively. Spontaneous TR symmetry breaking TIs, without magnetic field, are known as quantum anomalous Hall (QAH) insulator in 2D, or topological axion insulator in 3D.[24,25] The axion insulator has a quantized magnetoelectric response identical to that of a (strong) TI, but lacks the protected surface states of the TI. Other methods of obtaining insulating state such as disorder, Kondo effect, or Hubbard interaction, associated with odd number of band inversions, are also proposed to give topological Anderson insulator,[26,27] topological Kondo insulator,[28] and topological Mott insulator[29], respectively. Finally, a new class of TI is proposed by the present author, which is called quantum spin-Hall density wave (QSHDW) insulator in quasi-2D.[30] When two opposite chiral states are significantly nested in a given system, it renders a transitional symmetry breaking Landau order parameter (forming spin-orbit density wave[31,32]), which can be associated with odd number of band inversions and $Z_2$ topological invariant for a special nesting vector. In this case, the parity or chirality inversion occurs in the real space between different lattice sites, breaking the transitional symmetry, but not the TR symmetry.

Density functional theory (DFT) [18,19] calculations take a preceding role in predicting most of the TIs, many of which are followed by experimental realizations. Owing to weak SOC in graphene, this material has not been realized to be intrinsic TI despite its first prediction. HgTe/CdTe quantum well state were predicted to be 2D TI,[9] which was followed by its experimental realization.[33] The 3D topological semimetal predicted,[8] and realized[34] is $Bi_{1-x}Sb_x$. The first 3D TI was discovered in $Bi_2Se_3$, $Bi_2Te_3$, $Sb_2Te_3$, family both theoretically and experimentally.[18,19,35] This is followed by a series of predictions of 3D TIs including gray tin,[7] HgTe, InAs,[20] ternary tetramytes $Ge_mBi_{2n}Te_{(m+3n)}$ series,[36] half-Huesler compounds,[37,38] Tl-based III-V-VI2 chalcogenides,[39,40] ternary I-III-VI2 and II-IV-V2 chalcopyrites, I3–V–VI4 famatinites, and quaternary I2–II–IV–VI4 chalcogenides,[41] $Li_2AgSb$,[20] LiAuSe honeycomb lattice,[42] $β-Ag_2Te$,[43] non-centrosymmetric BiTe$X$ ($X$=Cl, I, Br)[44–46]. Recently a number of materials are discovered to have stable 2D structure, among which Si, Ge[46], Sn,[47] As,[48] Bi,[49,50] P[51] are predicted to exhibit QSH insulating state with SOC, and with other tuning. $Pb_{1-x}Sn_xSe/Te$[52–54] and SnS[55] are the only topological crystalline insulators known to date. *f*-electron based compounds such as $SmB_6$,[28,56] $YbB_6$[57] are predicted to be topological Kondo insulators, while $PuB_6$ is considered a topological Mott insulator.[58] $URu_2Si_2$ is considered as a candidate material for the spin-orbit density wave induced hidden topological order system.[31,59,60] Ir, and Os-based oxides are proposed to be axion insulators.[25] The list of QAH insulator is rather small,[61] most of which require external tuning with magnetic doping and thin films. La$X$ ($X$ = Br, I and Cl) family is predicted to be intrinsic QAH insulator with sizable band gap for applications.[62] A complete materials repository and their individual properties can be found in Refs. [63,64].

Despite this tremendous success as well as continuing research activities for harvesting more TI materials, the real struggle with the presently available materials is that the unwanted bulk conduction band lies below the Fermi level, and contributes to the transport phenomena. There have been considerable efforts to eliminate the bulk band above the Fermi level by chemical doping,[65,66] pressure,[67,68] photo-doping,[69] heterostructure,[70,71] etc. All these efforts have however little success and the ultimate aim for obtaining pure edge current remained unachieved. In the 2D counterpart, the challenge extends to the lack of material diversity. Moreover, although HgTe/CdTe[33] and InAs/GaSb[72] are demonstrated by transport measurement to show QSH behavior, but for other measurements, these samples are not very useful.

A jump-start to the field was offered by various engineering principles for TIs. There have also been considerable efforts to engineer TIs by proximity induced SOC in graphene through adatoms,[73] or with interfacing with transition metal di-chalcogenides[74,75], and in artificial heterostructures,[5,16,76] and in optical lattices.[77–79] Artificially grown heterostructures can be very versatile as they offer higher materials flexibility and tunability. The present author has proposed a number of design principles for engineering quasi-1D,[30] 2D,[79] as well as 3D TI[5,16] by decorating atoms or layers. The basic idea lies in generating 'chiral orbits' in a periodic fashion without breaking TR symmetry in 2D plane. Each 'chiral orbit' thereby encloses a pseudo-magnetic field (Berry curvature) in the same sense as in the TKNN language and commence quantized Chern number.[79] In real-space, electron-electron interaction induced translational symmetry breaking can lead to chirality inversion between different sublattices, leading to a new type of $Z_2$ quantum order which is associated with topological invariant and edge states.[30] The 3D generalization follows similarly in that the adjacent layers should have opposite SOC, such that they serve as TR partner to each other. As the quantum tunneling between them drives a band inversion, a 3D TI naturally arises.[5] A heterostructure of even and odd orbital orbitals is also a fertile setup to generate Dirac materials.[16] The second part of this article discusses various forms of these engineering principles.

TI has been reviewed extensively in various forms. Three review articles are published in the Review of Modern Physics.[23,63,80] Topological band theory[63,81] and topological field theories[14] are also discussed extensively. A materials repository can be found in Refs.[63,64] Topological superconductors are reviewed in Refs. [23,82,83]. QSH[84] and QAH[61] insulators are also reviewed separately. Reviews of Dirac and Weyl materials can be found in Refs. [85,86]. Few books are available to understand the basics of various TIs .[87,88]

## II. Theories of topological invariants

Topological band theory encompasses broadly defined computational schemes for the calculations of various topological invariants in systems with or without TR symmetry, inversion symmetry, particle-hole symmetry, and mirror symmetry. Depending on larger number of symmetries present in the Hamiltonian, the computation of the corresponding topological invariant simplifies accordingly. Both Chern number and $Z_2$ invariants are defined by single particle wavefunction, and can be calculated using the tight-binding or Wannier wavefunction within the DFT. For weakly interacting systems, such calculation can be extended to the quasipaprticle spectrum within a Fermi-liquid or mean-field theory. For strongly correlated systems, the Chern number can be calculated by using the self-energy dressed Green's function.[89,90] In addition to rigorous calculations of Chern number or $Z_2$ invariants, there are also other simplified methods which can be used for preliminary diagnosis of a potential TI. For examples, one can determine the non-trivial TI by simply counting the odd number of band inversions at the TR symmetric k-points, or employing the adiabatic (band gap-) continuity between a known TI and an unknown insulator, and/or bulk-boundary correspondence.

We start this section with the historical development of topology which forms the basis for modern topological invariant. Then we discuss various topological invariant calculations based on the number of symmetries present in a given case. Across most of the methods we discussed below, some unified concepts can be excavated which combine diverse formalisms of topological invariant. Among them, we have: (i) Complex phase associated with electron hopping with quantized phase winding; (ii) Formation of cyclotron or 'chiral orbits' in periodic lattice (Gaussian curvature); (iii) Odd number of exchange of electron's complex phase, or chirality or TR conjugate in half of the BZ; (iv) Search for bulk property (s) which can lead to accumulation of charge (IQH), or spin, (QSH), or TR polarizibility ($Z_2$ class). We will discuss below such corresponding basic principles used for the derivation of various topological invariants.

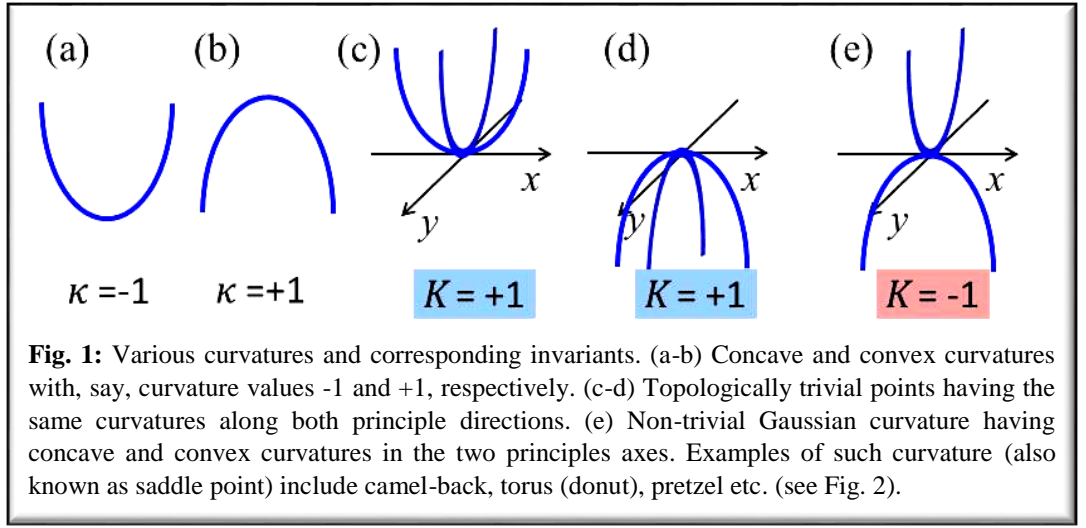

**Fig. 1:** Various curvatures and corresponding invariants. (a-b) Concave and convex curvatures with, say, curvature values -1 and +1, respectively. (c-d) Topologically trivial points having the same curvatures along both principle directions. (e) Non-trivial Gaussian curvature having concave and convex curvatures in the two principles axes. Examples of such curvature (also known as saddle point) include camel-back, torus (donut), pretzel etc. (see Fig. 2).

**A. Gaussian curvature, Euler number, and genus**

The inception of the concept of invariants can be traced back to the times of Leonhard Euler and Carl Friedrich Gauss in the 18th century. Gaussian curvature can be a good starting point to build up the discussion. It is defined by the product of two principle curvatures ($\kappa_1$, $\kappa_2$), along any two perpendicular directions at a given point (see Fig. 1) as $K = \kappa_1 \kappa_2$. If a principle curvature has a minimum (maximum) at the point (convex and concave curvatures), we assign its value to be -1 (+1). In this sense, if a point on the surface has minima or maxima in both principle directions, then the corresponding Gaussian curvature is +1 [Figs 1(c-d)]. The outer and inner surfaces of a sphere provide the corresponding examples, both being topologically equivalent. On the other hand, if a point simultaneously possess maximum and minimum in the two principle directions, the Gaussian curvature yields -1 [Fig. 1(e)]. The camel's back or a torus is non-trivial Gaussian curvature with $K = -1$.

Euler characteristic (also known as Euler number) dictates that the flux through a Gaussian curvature is always quantized as

$$\oiint_S K dS = 2\pi \nu, \text{ where } \nu = \text{integer.} \quad (1)$$

The above integral formula can also be expressed in terms of the Gauss-Bonnet formula, giving a topological invariant, called genus ($g$), relating the Euler characteristic as $\nu = 2 - 2g$. The Euler characteristic for a sphere is 2, giving $g = 0$ [Fig. 2(a)]. The same for a torus or Möbius strip is 0, with $g = 1$ [Fig. 2(b)]. Thus the former geometry is attributed as topologically trivial, while the later (torus, Möbius strip) as non-trivial curvature. Double torus and a three–hole pretzel have Euler characteristic as -2, and -4, with $g = 2$, and 3, respectively [Fig. 2(c)]. From these examples, it is evident that the genus or the topological invariant is related to the number of holes present in a Gaussian geometry. Another important observation can be made here that the number of holes or genus also dictates the number of distinct surface states. These results constitute the key mechanism for the emergence of topological invariants in the quantum and condensed matter world.

**B. Laughlin's argument and TKNN invariant**

In simple Hall effect, as the magnetic field is applied perpendicular to the lattice, a potential gradient arises perpendicular to both applied magnetic field and applied current. This is because, due to magnetic field, the electron and holes feel opposite Lorentz forces and move to different edges of the lattice (Fig. 3(b)). The transverse conductivity ($\sigma_{xy}$) initially increases linearly with the magnetic field strength (Fig. 3(a)). But with further increase of the field, the Hall conductivity becomes quantized and exhibits plateau with magnetic field, and increases only by integer multiple of $e^2/h$, $e$ and $h$ are the usual constants. This is the first realization[91] of topological invariant in physical systems.

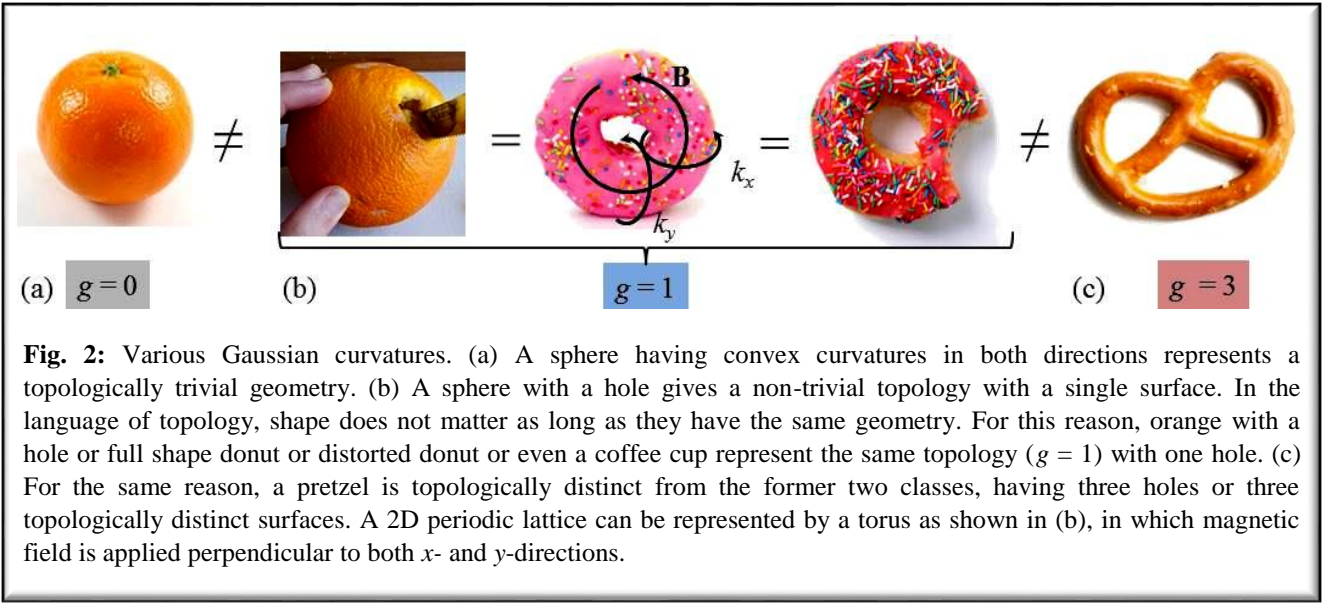

**Fig. 2:** Various Gaussian curvatures. (a) A sphere having convex curvatures in both directions represents a topologically trivial geometry. (b) A sphere with a hole gives a non-trivial topology with a single surface. In the language of topology, shape does not matter as long as they have the same geometry. For this reason, orange with a hole or full shape donut or distorted donut or even a coffee cup represent the same topology (*g* = 1) with one hole. (c) For the same reason, a pretzel is topologically distinct from the former two classes, having three holes or three topologically distinct surfaces. A 2D periodic lattice can be represented by a torus as shown in (b), in which magnetic field is applied perpendicular to both *x*- and *y*-directions.

In the QH regime, electrons form cyclotron orbits in the bulk and becomes localized (see Fig. 3(c)). Although magnetic field breaks translation asymmetry, but as the magnetic field is sufficiently large, the radius of the cyclotron orbits reduces. Here the cyclotron orbits form a larger magnetic unit cell, whose cross-sectional area changes with the field strength. R.B. Laughlin[1] recognized that the periodic lattice in a 2D plane can be represented by a torus, forming a non-trivial Gaussian curvature [Fig. 2(b)]. The magnetic flux through the magnetic torus is thus quantized, according to the Euler integral in Eq. (1), as

$$\varphi = \oiint_{S \in MT} B_z dS = B_z S = \nu\, h/e, \qquad (2)$$

where $B_z$ is the perpendicular component of the applied magnetic field, $S$ is the cross-sectional area of the magnetic torus (MT), $\nu$ is integer. So the question is as the magnetic field is continuously increased, how the area of the magnetic unit cell changes to respect the above quantization condition and how the bulk topology arises?

Thouless, Kohmoto, Nightingale, and Nijs (TKNN)[2] argued that the area of the magnetic unit cell increases as integer multiple of the original unit cell ($S_0 = ab$) as, $S = qS_0$, where $a$ and $b$ are the lattice parameters and $q$ is an integer. Therefore the flux through the original unit cell is a rational number times the flux quanta ($\varphi_0 = h/e$.): $\varphi = B(ab) = \frac{\nu}{q}\varphi_0$. Once a magnetic unit cell is defined, we can now Fourier transform to the corresponding momentum space by redefining a magnetic translational symmetry and quantify a bulk topological invariant. We recall that here the Hall conductivity is itself a topological invariant: $\sigma_{xy} = \nu e^2/h$, where $\nu$ is called TKNN or Chern number. Hall conductivity can be calculated from the Kubo formula using current-current correlation function. Therefore, from the general Kubo formula for conductivity, we can obtain our first definition of a topological invariant or the Chern number for the $n^{th}$ band ($\nu_n$) as

$$\nu_n = i \sum_{\mathbf{k}, n' \neq n} (f(E_{n\mathbf{k}}) - f(E_{n'\mathbf{k}}))$$

$$\times \frac{\left[\left\langle \psi_n(\mathbf{k}) \left| \frac{\partial H}{\partial k_x} \right| \psi_{n'}(\mathbf{k}) \right\rangle \left\langle \psi_{n'}(\mathbf{k}) \left| \frac{\partial H}{\partial k_y} \right| \psi_n(\mathbf{k}) \right\rangle\right]}{(E_{n\mathbf{k}} - E_{n'\mathbf{k}})^2} \qquad (3)$$

where $E_{nk}$ is the eigenvalue of the Hamiltonian $H$ and $\psi_n(k)$ is the Wannier wavefunction, and $f(E_{nk})$ is the Fermi-Dirac distribution function.

### C. Quantum Hall calculation in arbitrary parameter space

The above formula is based on the variation of the Hamiltonian in two orthogonal momentum directions and thus implicitly assumes a periodicity of the lattice. In some cases, as in disordered lattice, where a proper unit cell is difficult to define, the above formula apparently fails. However, Niu, Thouless and Wu[92] generalized their TKNN invariant calculations to any two arbitrary parameter space, by implementing the so-called "twisted boundary condition". They introduced two fictitious parameters $\alpha, \beta$ (which does not require to have any physical relevance) and demanded that the lattice is periodic under them as

$$\psi(x_i + L_1) = e^{i\alpha L_1} e^{i(eB/\hbar)y_i L_1}\psi(x_i),$$
$$\psi(y_i + L_2) = e^{i\beta L_2}\psi(y_i). \qquad (4)$$

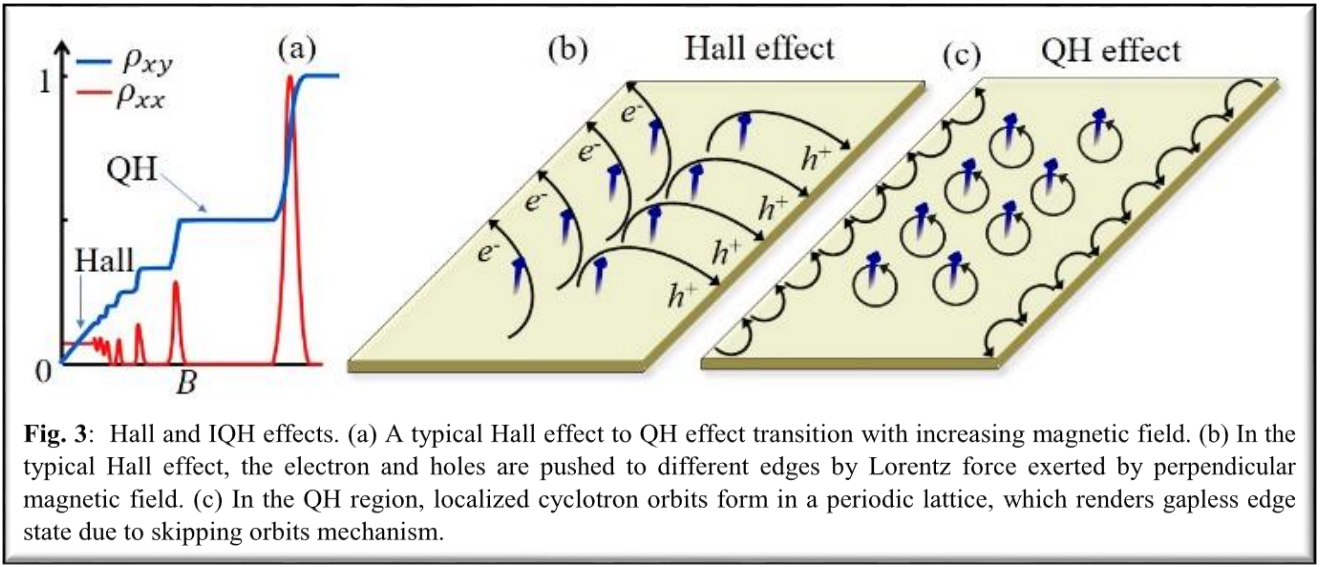

**Fig. 3**: Hall and IQH effects. (a) A typical Hall effect to QH effect transition with increasing magnetic field. (b) In the typical Hall effect, the electron and holes are pushed to different edges by Lorentz force exerted by perpendicular magnetic field. (c) In the QH region, localized cyclotron orbits form in a periodic lattice, which renders gapless edge state due to skipping orbits mechanism.

Here $x_i, y_i$ are the lattice site indices, and $L_1, L_2$ represent the system size. In this case, the velocity operators can be rewritten as $v_x = \frac{\partial H}{\partial k_x} \to \frac{\partial H}{\partial \alpha}$, and $v_y = \frac{\partial H}{\partial k_y} \to \frac{\partial H}{\partial \beta}$, and the QH invariant can be calculated in the $(\alpha, \beta)$ space by using Eq. (3). This generalization unravels an important insight that as the Hamiltonian is adiabatically varied in any closed parameter space, it gives the same topological invariant. This implies that as the particle returns to its starting points, the expectation value of the velocity operators remains the same, but its wavefunction itself acquires an additional phase. This phase turns out be the Berry phase, proposed independently.[93] The expectation value of the velocity operator in any parameter space is an important factor for topological invariant, with the only requirement for a given parameter space is that it has to be periodic (Gaussian curvature), but not necessarily a physical parameter. This is crucial for the conceptualization of the 'chiral orbit' we use for the QSH effect which implies that 'chiral orbit' can be a mathematical object which can be 'created' in any periodic parameter space for the calculation of the topological invariant in 2D systems. For the same reason, when a system is driven periodically with time, the corresponding time evolution of the Hamiltonian (Floquet Hamiltonian) can also give rise to a 'Berry phase' in the time-domain and lead to QH or topological phase.

### D. Berry connection and curvature

The above section discussed how an applied magnetic field's flux quantization leads to the IQH as computed within the Kubo formula. Now we can reverse our derivation, and start with the Kubo formula version of the topological invariant in Eq. (3), and define a band dependent 'magnetic field' in the momentum space $\mathbf{F}_n(\mathbf{k})$. We again demand that its flux in the reciprocal space is quantized:

$$\nu_n = \frac{1}{\Omega_{BZ}} \iint_{BZ} \mathbf{F}_n(\mathbf{k}) \cdot \mathbf{n}\, d^2k = \oint_{\partial BZ} \mathbf{A}(\mathbf{k}) \cdot d\mathbf{k}, \quad (5)$$

where $\Omega_{BZ}$ is the BZ phase space area. In the last step, we have employed the Stokes' theorem, which allowed us to define a momentum space 'vector potential' as $\mathbf{F}_n(\mathbf{k}) = \nabla_k \times \mathbf{A}_n(\mathbf{k})$. The formalism for $\mathbf{F}(\mathbf{k})$ is simply the right-hand side of Eq. (3), and that for $\mathbf{A}(\mathbf{k})$ can also be obtained subsequently. Another elegant formalism for $\mathbf{F}$ and $\mathbf{A}$ can be obtained by using the identity

$$|\partial_{k_i}\psi_n(\mathbf{k})\rangle = \sum_{n'\neq n} \frac{\langle \psi_{n'}(\mathbf{k})| \frac{\partial H}{\partial k_i} |\psi_n(\mathbf{k})\rangle}{E_{n\mathbf{k}} - E_{n'\mathbf{k}}} |\psi_{n'}(\mathbf{k})\rangle, \qquad (6)$$

which yields

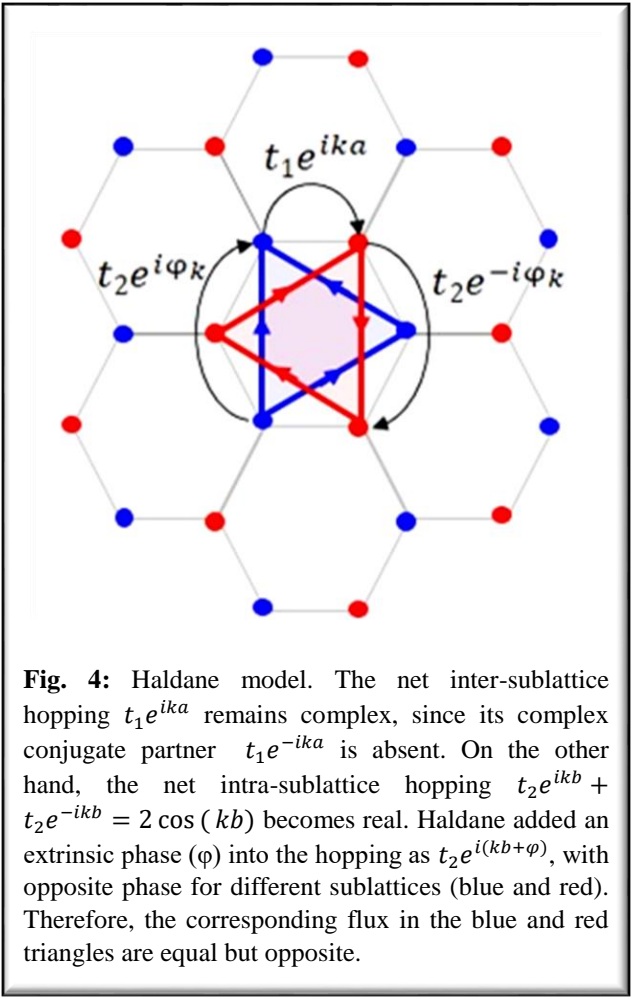

**Fig. 4:** Haldane model. The net inter-sublattice hopping $t_1 e^{ika}$ remains complex, since its complex conjugate partner $t_1 e^{-ika}$ is absent. On the other hand, the net intra-sublattice hopping $t_2 e^{ikb} + t_2 e^{-ikb} = 2\cos(kb)$ becomes real. Haldane added an extrinsic phase (φ) into the hopping as $t_2 e^{i(kb+\varphi)}$, with opposite phase for different sublattices (blue and red). Therefore, the corresponding flux in the blue and red triangles are equal but opposite.

$$\mathbf{F}_n(\mathbf{k}) = i\langle\nabla_k\psi_n(\mathbf{k})| \times |\nabla_k\psi_n(\mathbf{k})\rangle, \qquad (7)$$

$$\mathbf{A}_n(\mathbf{k}) = i\langle\psi_n(\mathbf{k})|\nabla_k\psi_n(\mathbf{k})\rangle. \qquad (8)$$

To further elucidate the physical significance, we refer back to Eq. (5) which can be can be compared with the Peierls phase in real space, acquired by a charged particle moving in a magnetic field, $\varphi = \int_{c_1}^{c_2} \mathbf{A}(\mathbf{r}) \cdot d\mathbf{l}$, where $\mathbf{A}(\mathbf{r})$ is the vector potential, and $c_1$ and $c_2$ are starting and end points of the path. This implies that the topological invariant here is a momentum space 'Peierls phase' (equivalent to Aharonov-Bohm phase) acquired by the electron in traversing a closed path in the reciprocal space under an intrinsic gauge field $\mathbf{A}(\mathbf{k})$. In this sense, $\nu_n$ is called the Berry phase,[93] and $\mathbf{A}(\mathbf{k})$ as the Berry connection, while $\mathbf{F}(\mathbf{k})$ is the Berry curvature. Note that the Berry connection is gauge-dependent and therefore topological invariant formulas [such as axion angle formalism in Eq. (19) below] involving it does not give an unambiguous result. On the other hand, the Berry curvature is gauge invariant and observable. Therefore, for a band which possess a well-defined Berry phase in a close trajectory in the momentum space (translational symmetry is assumed as above), it possess an intrinsic Chern number, and therefore can give rise to an IQH effect without the application of an external magnetic field.

In the IQH effect, applied magnetic field provides the 'chirality' for the electrons to form cyclotron orbit. Without magnetic field, the QH phenomena can be thought of occurring in a reverse fashion. Here a self-generated chirality of electrons creates a pseudo-magnetic field (Berry curvature) in the process of forming 'chiral orbits'. In solid state systems, such intrinsic chirality can stem from a multiple origins, including SSH type staggered electron hopping,[15] sublattice (as often referred to pseudospin) symmetry in the hexagonal lattice,[94] SOC,[3] or certain type of even-odd orbital texture mixing[16,17]. In a simpler term, the chirality arises if the electron hopping is complex, because it naturally accompanies a phase associated with electron's hopping. As the $k$-space magnetic field or the Berry curvature threads through a periodic lattice (Gaussian curvature), the Euler characteristic ensures a quantization of the flux [Eq. (1)], and bulk topological invariant arises. The intrinsic formation of 'chiral orbit' in a periodic lattice is the foundation of TR invariant TIs, which however have different interpretations and mathematical expositions such as 'Pfaffian nodes', 'chiral vortex', momentum-space monopoles etc. as we will uncover below.

**E. Chern number without magnetic field**

If an 1D chain is made of two inequivalent sublattices, the hopping between the two sublattices becomes complex as used in the SSH model[15] According to the above prescription, the emergent 'chiral' hopping can be associated with a winding number. Recall that in a QH state, since the cyclotron orbits cannot complete a full circle at the edges, it leads to the edge current, and thus Hall effect arises. Something similar happens in 1D chain with chiral state. Since a localized chiral orbit cannot be assumed in 1D, the chirality of the electrons can be thought of as charge current flowing across the chain, allowing electrons and hole to be accumulated in opposite ends. This phenomena naturally arises if we solve Eq. (5) with open boundary conduction, which gives topologically protected polarizibility at the ends. This is called the Zak phase,[95] which is a topological invariant (discussed further in Sec. IIM below).

F.D.H Haldane[3] realized that such complex hopping can be easily obtained in 2D honeycomb lattice for the same reason, namely, due to the presence of two inequivalent sublattices. Two sublattices form triangular lattices, which are oppositely aligned, see Fig. 4. The low-energy Hamiltonian of a honeycomb lattice can be written in terms of the $2 \times 2$ Pauli matrices ($\sigma$) entangled with linear momentum, in which two sublattices provide the pseudospin spinor basis:[94]

$$H(\mathbf{k}) = \mathbf{d}(\mathbf{k}).\boldsymbol{\sigma}, \quad (9)$$

where $d_{1,2}(\mathbf{k})$ contain linear-in-$k$ term, and $d_3$ gives the Dirac mass. Such Hamiltonian is analogous to the Dirac equation in 2D, and forms Dirac cone in the absence of the Dirac mass term. For such simplified Hamiltonian, the Chern number can be calculated from the **d**-vectors itself. Starting from Eq. (3) and substituting (9), we obtain

$$\nu = \frac{1}{4\pi}\int_{BZ} d^2\mathbf{k}\, \frac{\mathbf{d}.\left(\partial_{k_x}\mathbf{d} \times \partial_{k_y}\mathbf{d}\right)}{|\mathbf{d}|^3}. \quad (10)$$

For the usual Dirac Hamiltonian, $d_i$ components are proportional to $k_i$ (where $i$ = x, y), and therefore, it is easy to see that the Chern number is proportional to the $d_3$ term. $d_3$ term stems from the onsite energy difference between the two sublattices, and intrinsically remains zero. A key ingredient is still missing here. Note that honeycomb lattice provides an imaginary hopping term between different sublattices, but the intra-sublattice hopping still remains real [Fig. 4]. Therefore, electron hopping within each triangular sublattice does not have any chirality and fails to forms our desired 'chiral orbit'. For a remedy, Haldane affixed an 'extrinsic' gauge field (but not a magnetic field) to the intra-sublattice hopping. Additionally, he imposed the condition that the 'gauge field' has different signs for different sublattices, such that the resulting 'chiral orbits' for them are counter-propagating. Therefore, they tread opposite flux and the net magnetic field effect remains zero. However, the staggered 'gauge field' naturally induces different onsite energy to different sublattices, and therefore, $d_3$ term becomes finite, and Eq. (10) gives a finite Chern number. This signifies that two oppositely rotating triangular 'chiral orbits' are split by a negative Dirac mass. Haldane's proposal was important for the conceptual development of the QH effect without magnetic field, but for decades, it was assumed to be 'unphysical' since obtaining the required 'gauge field' without magnetic field was not feasible. Very recently, researchers have successfully generated Haldane model by commencing time-dependent Hamiltonian with periodic pumping. The periodic time evolution naturally gives a 'Bloch phase' in the time-space which provides Haldane's 'gauge-field'.[96]

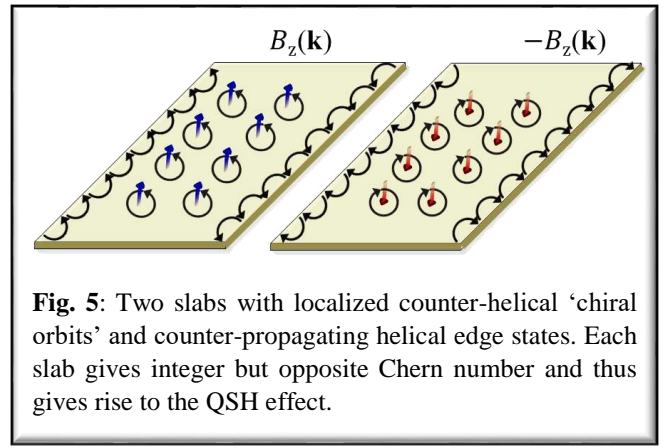

**Fig. 5**: Two slabs with localized counter-helical 'chiral orbits' and counter-propagating helical edge states. Each slab gives integer but opposite Chern number and thus gives rise to the QSH effect.

**E1. Spin Chern number**

Kane and Mele turned on SOC to obtain Haldane's 'gauge-field', which does not break TR symmetry.[4] This gave birth to the TR invariant QH effect and eventually $Z_2$ TIs. They considered two copies of Haldane's Hamiltonians for spin-up and spin-down states to form a Block diagonal Hamiltonian: $H(k) = h_\uparrow(k) \oplus h_\downarrow(k)$. The Hamiltonian respects both TR and inversion symmetry with $h_\uparrow(k) = -h_\downarrow^*(-k)$. The resulting Hamiltonian can be expressed in the $4 \times 4$ Dirac matrix basis as $H(\mathbf{k}) = \mathbf{d}(\mathbf{k}).\boldsymbol{\Gamma}$, where **d**-vector has five components and the corresponding $\boldsymbol{\Gamma}$ (including identity matrix) are the usual Dirac matrices (Kane and Mele used few additional cross-terms which we do not discuss here for simplicity). The $k$-dependence of each $d_i$ component ($d_0$ and $d_4$ contain even power of $k$, while others contain odd power) complements the symmetry of their corresponding $\Gamma_i$ matrices to preserve the TR symmetry. In the case when two valence bands are fully spin-polarized, the Chern number for each band corresponds to different spin states ($\nu_\uparrow, \nu_\downarrow$). Due to SU(2) symmetry $\nu_\uparrow = -\nu_\downarrow$, which means two equal and opposite 'chiral orbits' are stabilized due to the spin-momentum locking, as shown in Fig. 5. Therefore the total Chern number $\nu = \nu_\uparrow + \nu_\downarrow$ vanishes, while their difference, namely spin Chern number, $\nu_s = \nu_\uparrow - \nu_\downarrow$, become finite. Therefore, according to Kane and Mele, there is no charge pumping to the edge, but there is a net spin pumping, and hence they refer the corresponding state as QSH effect.

Bernevig, Hughes, Zhang (BHZ) predicted that the quantum well (QW) states arising in the HgTe/CdTe heterostructure commence 2D QSH insulator above some critical thickness.[9] They realized that the band structure for HgTe and CdTe are completely inverted across the Fermi level. In particular, the SOC split bands with $\Gamma_8$ and $\Gamma_6$ symmetries, respectively, constitute conduction and valence bands in HgTe, while they form valence and conduction bands in CdTe. Therefore, if we derive a 2 label Hamiltonian and expresses in terms of Pauli matrices as in Eq. (9), we immediately find that the Dirac mass $d_3 < 0$ for HgTe and $d_3 > 0$ for CdTe. Therefore, if one makes an HgTe/CdTe heterostructure, at their boundary $d_3$ must vanish, which means gapless Dirac fermions emerge here. Based on this idea, they proposed a $4 \times 4$ Hamiltonian using the Kramers pairs of $\Gamma_8$ and $\Gamma_6$ levels. The Hamiltonian is also block diagonal with each block representing different spin state as in the Kane-Mele model. The Dirac mass inversion (which is same as band inversion or parity inversion as we will discuss in Sec. IIG below) guarantees that each block gives equal but opposite Chern number, and QSH insulator arises. Due to the block diagonal nature of the mode, it is popularly known as half-BHZ model.

When two spin states cannot be separated to assign individual Chern number, the present method does not work. Kane and Mele proposed more rigorous method to calculate the $Z_2$ invariant using TR 'polarization' which is discussed in Sec. IIF. As we will go along, we will learn more techniques and interpretations of various topological invariances.

### E2. Mirror Chern number

In the cases, where the band inversion occurs at non-TR symmetric points, a distinct topological invariant can be obtained if the system possess mirror symmetry.[8,21,53] Let us consider a case where $\mathbf{k}_m$ represents a mirror plane in the BZ with the corresponding mirror operator ($\mathcal{M}$) defined by $[H(\mathbf{k}_m), \mathcal{M}] = 0$. In such a case, the mirror plane can be decomposed into two subspaces, denoted by $\pm \mathcal{M}$. Then, as in the case of half-BHZ model, the present Hamiltonian on the mirror plane can be split into two blocks, coming from two sub-space as $H(\mathbf{k}_m) = h_{+m}(\mathbf{k}_m) \oplus h_{-m}(\mathbf{k}_m)$. Each block ($h_{\pm m}$) gives equal but opposite Chern number (due to TR symmetry). Therefore, their difference $\nu_m = (\nu_{+m} - \nu_{-m})/2$ leads to a finite value, called mirror Chern number. The corresponding TI family is refereed as topological crystalline insulator.[21]

Given that the low-energy Hamiltonian here has mirror symmetry, the leading term in the edge state will have even power in momentum along this direction. Let us consider an example of a mirror plane at $k_x = 0$, which dictates $H(k_x, k_y, k_z) = H(-k_x, k_y, k_z)$. Since a linear term in $k_x$ violates this condition, it will drop out from the Hamiltonian. Therefore, the corresponding surface state will be quadratic. Along other directions lacking a mirror symmetry, the edge/surface state can be linear in momentum. In fact, the surface states of the topological crystalline insulator $Sn_{1-x}Pb_xTe$.[81] contain both quadratic and linear bands.

### F. $Z_2$ invariant and time-reversal polarization

In the case of TR breaking IQH insulator, Chern number can take any arbitrary value. However, this is not the case for TR invariant TIs. For such cases, spin or mirror Chern number can take only 0 or 1 (mod 2) value, and thus the topological invariant is represented by a more general $Z_2$ invariant.[4,6] $Z_2$ invariant becomes equal to spin or mirror Chern number in the cases the latter are defined, but there exists other methods of evaluating it. Although spin and mirror Chern numbers are observables via QH effect, $Z_2$ invariant is not a directly measurable quantity in the bulk, and is often diagnosed by the observation of topological surface state.

For TR symmetric cases, Kramers degeneracy at the TR invariant k-points dramatically reduces the full momentum space calculations into only TR invariant k-points. The antiunitary TR operator $\Theta$ imposes the symmetry in the Hamiltonian as $\Theta H^*(\mathbf{k}) \Theta^{-1} = H(-\mathbf{k})$. Let us consider a system where the spin-rotational symmetry is broken, say due to SOC, without breaking the TR symmetry as $\Theta |\mathbf{k}, \uparrow\rangle =$ $|-\mathbf{k}, \downarrow\rangle$. The high-symmetric points $\mathbf{k}_s^*$, which are invariant under TR symmetry [e.g. (0,0,0), ($\pi$,0,0), (0,$\pi$,0), ($\pi$, $\pi$,0), ($\pi$, $\pi$, $\pi$), etc in a cubic lattice in Fig. 6] are special. Here $\mathbf{k}_s^*$ and $-\mathbf{k}_s^*$ points are the same [up to a reciprocal lattice vector: $\mathbf{k}_s^* = -\mathbf{k}_s^* + \mathbf{G}$], so they must be spin degenerate. This is called the Kramers' degeneracy.

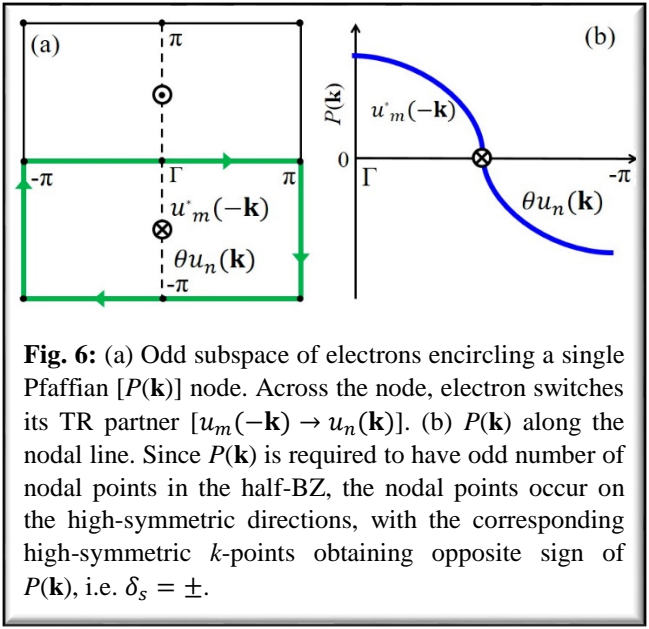

**Fig. 6:** (a) Odd subspace of electrons encircling a single Pfaffian [$P(\mathbf{k})$] node. Across the node, electron switches its TR partner [$u_m(-\mathbf{k}) \rightarrow u_n(\mathbf{k})$]. (b) $P(\mathbf{k})$ along the nodal line. Since $P(\mathbf{k})$ is required to have odd number of nodal points in the half-BZ, the nodal points occur on the high-symmetric directions, with the corresponding high-symmetric $k$-points obtaining opposite sign of $P(\mathbf{k})$, i.e. $\delta_s = \pm$.

Among many methods available for the evaluation of the $Z_2$ invariant, Fu-Kane-Mele method is often easier to implement, especially in the cases where both TR and inversion symmetries are present.[7] To understand this method, we draw analogy with some of the properties of IQH, QSH states discussed above. In these insulators, it is the bulk Chern number which induces charge or spin polarizations, respectively, at the edge. Kane and Mele asked a similar question: what bulk property for TR invariant $Z_2$ class can pump a similar 'polarization' to the boundary. In QSH effect, opposite spins with opposite momentum, due to SOC, are pumped to the edge, requiring that the electron exchanges its spin in traversing half of the BZ odd number of times. Since opposite spins with opposite momentum are just the TR conjugate to each other, a more fundamental property to exchange in $Z_2$ TI is the TR partner of electrons. Based on this analogy, Kane, Mele proposed a mathematical concept, called 'TR polarization', in which they argued that electrons with one Bloch wavefunction and their complex conjugate partner are accumulated at the edge.[4,97] This requires that the electron switches its TR partner odd number of times in traversing half of the BZ [green line in Fig. 4(a)]. [For systems with inversion symmetry, it is equivalent to the odd number of parity, or equivalently the Dirac mass or just simply band inversion in half of the BZ, as discussed in Sec. IIG.]

They subsequently quantified this hypothesis[4,6,7,97] by defining the matrix element of the TI operator between a Bloch state $u_n(\mathbf{k})$ and its TR conjugate $u_m^*(-\mathbf{k})$ in the Fermi sea, and construct an antisymmetric, unitary matrix, with components $w_{mn}(\mathbf{k}) = \langle u_m(-\mathbf{k})|\Theta|u_n(\mathbf{k})\rangle$. The determinant of the antisymmetric matrix $w$ is represented by the Pfaffian as $[Pf(w)]^2 = \det(w)$. We define $P(k) = \text{Pf}[w(k)]$. For many TI Hamiltonians dealing with SU(2) spin, the filled state is two-fold degenerate, especially when inversion symmetry is present. Therefore, the above matrix-element is a $2 \times 2$ matrix in which the Pfaffian is just the off-diagonal term (the formula for topological invariant is, however, general to any number of filled bands).

Based on the value of $P(\mathbf{k})$, the BZ can be split into the 'even' and 'odd' subspaces. In the even subspace, $\Theta|u_n(\mathbf{k})\rangle$ is proportional to $|u_m(-\mathbf{k})\rangle$, making $|P(\mathbf{k})| = 1$. In the odd subspace, $\Theta|u_n(\mathbf{k})\rangle$ is orthogonal to $|u_m(-\mathbf{k})\rangle$, implying $P(\mathbf{k}) = 0$, which is important for topological invariance. Let us assume that $\pm\mathbf{k}^*$ is a pair of k-points, which are TR partners, where $P(\mathbf{k}^*) = 0$ [see Fig. 4(a)]. The phase of $P(\mathbf{k})$ about each of these points winds in opposite directions (equivalent to having counter propagating 'chiral orbits' in the momentum space). If $\pm\mathbf{k}^*$ coincides with any TR invariant point $\mathbf{k}_s^*$, the two $k$-space 'chiral orbits' annihilate each other. Again if there are even number of such points, say, $\pm\mathbf{k}_{1,2}^*$, then unless $\mathbf{k}_{1,2}^*$ are protected by some additional symmetry, they can also annihilate each other by scattering or perturbation. But a single pair of $\pm\mathbf{k}^*$ does not have the option to scatter to another $k$-point, except to the corresponding $\mp\mathbf{k}$-points, which however requires the corresponding spin to flip. Since spin flip is prohibited by the TR symmetry, such nodal points at $\pm\mathbf{k}^*$ remain protected from TR invariant perturbations.

Similarity between the Berry connection formalism, and the Pfaffian $P(\mathbf{k})$ can be rigorously shown. Differentiating $w(\mathbf{k})$, and using the unitary property of the $w$ matrix, we obtain the Berry connection in terms of $w(\mathbf{k})$, and $P(\mathbf{k})$ as[7] [using Eq. (8)]

$$\mathbf{A}(\mathbf{k}) = -\frac{i}{2}\text{Tr}[w(\mathbf{k})^\dagger \nabla_\mathbf{k} w(\mathbf{k})] = -\frac{i}{2}\text{Tr}[\nabla_\mathbf{k} \log w(\mathbf{k})]$$

$$= -\frac{i}{2}\nabla_\mathbf{k} \log \det[w(\mathbf{k})] = -i\nabla_\mathbf{k} \log[P(\mathbf{k})] \quad (11)$$

Based on this, $Z_2$ invariant can be defined by calculating the winding number of the $P(\mathbf{k})$ over a single **k**-space 'chiral orbit', in a contour enclosing half of the BZ (so that only either $+\mathbf{k}^*$ or $-\mathbf{k}^*$ is included), as shown by green boundary in Fig. 4(a). So, from Eq. (5), $Z_2$ invariant can be evaluated as:

$$\nu = \frac{1}{2\pi i}\oint_C d\mathbf{k} \cdot \nabla_\mathbf{k} \log[P(\mathbf{k}) + i\delta], \quad (12)$$

where a complex term $i\delta$ is introduced to evaluate the above integral in a complex contour plane. This simplifies the integral into a residue problem, with singularities occurring at the loci of $P(\mathbf{k}^*) = 0$.

Given that there should be an odd number of Pfaffian nodes in half of the BZ, it is expected that the corresponding nodes would occur on the high-symmetric $k$-directions. This simply means that $P(\mathbf{k})$ should change sign odd number of times on both sides of the nodes at the TR invariant **k**-points [see Fig. 4(b)]. Therefore, the calculation simply reduces to counting the sign of $P(\mathbf{k})$ at the TR invariant momenta in the first quadrant of the BZ only. If we take the product of the sign of the Pfaffian at all TR invariant points, and the result comes out to be negative, then there must be odd number of zeros in the first quadrant of the BZ. Since $[\text{Pf}(w)]^2 = \det(w)$, the sign of the Pfaffian can be defined in a formal way as

$$\delta_s = \frac{\sqrt{\det[w(\mathbf{k}_s^*)]}}{\text{Pf}[w(\mathbf{k}_s^*)]} = \pm 1, \quad (13)$$

Therefore, in a 1D system, the TR polarization can be defined as $(-1)^\nu = \delta_1 \delta_2$, where $\delta_1$, and $\delta_2$ are evaluated at the two TR invariant points. If $\delta_1$, and $\delta_2$ have opposite sign, we get the $Z_2$ invariant $\nu = 1$, which signals the non-trivial topological phase. The formula generalizes to higher dimensions as

$$(-1)^\nu = \prod_{s=1}^{N_s} \delta_s, \quad (14)$$

where $N_s$ is the total number of the TR invariant momenta in the first quadrant of the BZ. In a 2D square lattice, $N_s = 4$, while in a 3D $C_4$ symmetric lattice $N_s = 8$. If there are odd number of $\delta_s = -1$ in this $k$-space, the right hand side of the above equation gives -1, which therefore yields $\nu = 1$, a non-trivial topological invariant. This is called the strong topological invariant (denoted by $\nu_0$). Note that for any arbitrarily large odd number of $\delta_s = -1$, topological invariant remains $\nu_0 = 1$, otherwise 0. Therefore, unlike in IQH insulator where arbitrarily large Chern number is possible, here one only gets two values of $\nu_0$ and the $Z_2$ symmetry emerges.

In some cases, there can be total even number of $k_s^*$-points with $\delta_s = -1$, but they lie in different planes (say on the $k_x = 0$, and $k_x = \pi$ planes) [see Fig. 5(c)]. Thus those 2D planes contain odd number of $\delta_s = -1$, and constitute non-trivial 2D TIs, while the 3D system remains trivial TI. This is called the weak TI. Since in 3D, there are three orthogonal coordinate axes, there are three weak topological invariants $(\nu_1, \nu_2, \nu_3)$. Fu, Kane, and Mele,[6] thereby, introduced four $Z_2$ invariants $(\nu_0: \nu_1 \nu_2 \nu_3)$ for 3D TIs. This part is explained with examples in Fig. 7 and discussed further in the following section.

**G. $Z_2$ calculation *with* inversion symmetry**

If the system possesses inversion symmetry, in addition to TR symmetry, calculation of topological invariants becomes exceptionally simpler. Suppose $\mathcal{P}$ is the parity operator defined by $\mathcal{P}|k,\uparrow\rangle = |-k,\uparrow\rangle$, under which the Hamiltonian transforms as $H(-\mathbf{k}) = \mathcal{P}H(\mathbf{k})\mathcal{P}^{-1}$. Inserting $\mathcal{P}^2 = 1$ in the expression for $w_{mn}(\mathbf{k})$, and employing the identity that $[H, \mathcal{P}\Theta] = 0$, $\delta_s$ parameter at the TR invariant moment $\mathbf{k}_s^*$ can be evaluated as[7]

$$\delta_s = \prod_{m=1}^{N} \xi_m(\mathbf{k}_s^*), \quad (15)$$

where $\xi_m(\mathbf{k}_s^*) = \pm 1$ is the parity eigenvalue at $\mathbf{k}_s^*$ defined as $\mathcal{P}|u_m(\mathbf{k}_s^*)\rangle = \xi_m(\mathbf{k}_s^*)|u_m(\mathbf{k}_s^*)\rangle$. The product is computed for $N$ filled bands. In practice one does not have to include all filled bands in the calculation, rather only those bands which undergoes band inversion at the TR invariant points. For the two typical Dirac Hamiltonians which are expressed in terms of $2 \times 2$ Pauli matrices or $4 \times 4$ Dirac matrices, the parity term turns out to be $\sigma_z$, or $\Gamma_4 = \sigma_z \otimes I_{2\times 2}$, respectively. The corresponding $k$-dependent **d**-vector component can often be written as $d_4(\mathbf{k}_s^*) = (\varepsilon_1(\mathbf{k}_s^*) - \varepsilon_2(\mathbf{k}_s^*))/2$, where $\varepsilon_{1,2}(\mathbf{k}_s^*)$ are the conduction and valence bands near the Fermi level. Therefore, the parity eigenvalue of the Hamiltonian is determined simply by $\xi_m(\mathbf{k}_s^*) = \text{sgn}[d_4(\mathbf{k}_s^*)]$. For

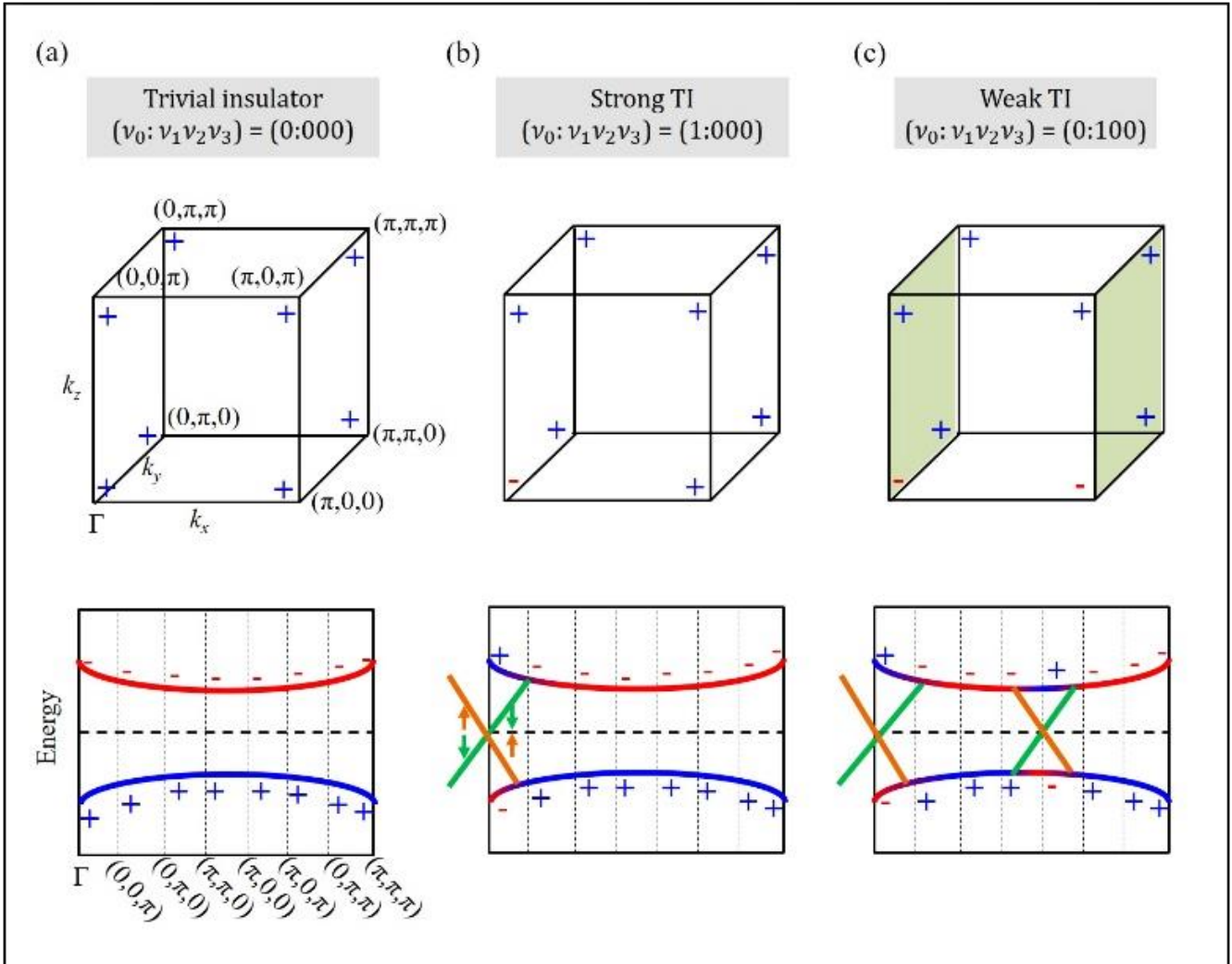

**Fig. 7**: Band inversion and non-trivial TI. According to Eqs. (14-15), topological invariant is determined by the total odd number of inversions of the Pfaffian sign or parity ($\delta_s$) in the *valence* band. In this figure, the choice of signs of $\delta_s$ for conducting and valence bands and the cubic lattice structure are completely arbitrary and for illustration purpose only. (a) There is no band inversion and thus it's a trivial band insulator. Even when the conducting band drops below the Fermi level at all k-points, the system still remains topologically trivial. (b) When the band is inverted at odd number of TR invariant **k**-points, one obtains strong TI ($\nu_0 = 1$). Each *k*-point where bands are inverted, a metallic Dirac cone arises at the corresponding edge or surface. (c) When even number of band inversion occurs, it does not render a *strong* 3D TI ($\nu_0 = 0$). But those 2D planes accommodating odd number of band inversions gives 2D TI (such as $k_y = 0$ and $k_y = \pi$ planes in this example): $(\nu_0: \nu_1\nu_2\nu_3)=(0:100)$.

systems with both inversion and TR symmetries, the $Z_2$ invariant is obtained by simply counting the number of band inversion at the high-symmetric momenta as:

$$(-1)^\nu = \prod_{s=1}^{N_s} \text{sgn}[\varepsilon_1(\mathbf{k}_s^*) - \varepsilon_2(\mathbf{k}_s^*)]. \qquad (16)$$

In other words, if there are odd number of band inversions at all TR invariant momenta in an insulator, the system acquires non-trivial topological behavior with the same odd pair of edge or surface states. Interestingly, $d_4$ term is the mass term in the Dirac equation. That means, parity inversion is equivalent to Dirac mass inversion from positive (trivial) to negative (non-trivial) value. For systems, where parity is not a good quantum number, the Dirac mass inversion serves as a good measure of the topological phase transition as often used in the literature (see Fig. 7).

The reason for the popularity of this method lies in the fact that identifying the parity of a given band within the tight binding model or Wannier method is quite straightforward. In most cases, it is nothing but knowing the orbital character of the valence and conduction bands. So band inversion simply refers to switching orbital character between these two classes.[16,17,81] However, band inversion does not mean that an orbital entirely switches its position between conduction and valence bands at all *k*-points, rather it has to be done only at odd number of TR *k*-points, and not at other k-points. For example, in Fig. 7(a), if the odd parity conduction band drops *fully* below the Fermi level, the system still remains topologically trivial. This means the inter-orbital overlap matrix-element has to be strongly momentum dependent. Simple local inter-orbital hopping or crystal field splitting or onsite interactions such as Hubbard *U* or Hund's coupling are often not adequate to commence such a **k**-dependent band inversion. Spin-momentum locking due to SOC does this job in most of the known TIs.

Some caution has to be taken for the cases when a band at a given TR invariant k-point is not fully orbitally polarized, rather it contains a mixture of both even and odd orbitals. In such cases, the band inversion mechanism cannot be considered as conclusive. In this context, a term called band inversion strength is often used which measures the amount of orbital weight is exchanged between the conduction and valence bands. Band inversion strength is also used as a measure the Dirac gap at the TR *k*-points [see, for example, Ref. [20]]. Associated with the orbital weight transfer, the band topology also changes in this process. For example, if the top of the valence band has an upward curvature, it changes to a downward curvature around the TR k-point after the band inversion. This structure is sometimes referred as 'dent' in the band structure, which is seen in the DFT band structure, as well in the experimental data.[5,65]

Three representative examples for trivial, strong, and weak TIs are given in Fig. 7. Owing to TR symmetry, it is sufficient to consider only the first quadrant of the BZ to count the number of band inversions, since the other **k**-points are related to them by TR symmetry. As mentioned earlier, if the conduction and valence bands possess the same parity at all *k*-points, but different among them, Eq. (15) suggests that it is a trivial topological insulator, or not a topological insulator [Fig. 7(a)]. Fig. 7(b) depicts the case of a single band inversion at the Γ-point, indicating a strong TI ($\nu_0 = 1$). In this case, all three surfaces of the lattice possess Dirac cones with the vertex of the cone lying at the same **k**-point where the band inversion has occurred. If the band inversions occur even number of times in the first quadrant, a *weak* TI can be obtained if one or more BZ sides possess odd number of band inversions. For example, in Fig. 7(c), we consider the case of two band inversions at the Γ- and at (π,0,0)-points, yielding $\nu_0 = 0$. But the $k_x = 0$ and $k_x = \pi$-planes contain only single band inversion, and the corresponding topological invariant becomes $\nu_1 = 1$, while $\nu_{2,3} = 0$. A weak 3D TI can be thought of a stacking of 2D TIs, each having edge states. Since there are even number of Dirac cones here, scattering between them due to impurity or correlation can open a gap, and thus they are not topologically protected. Thus this state is refereed as *weak* topological insulator.

Each TR invariant **k**-point possessing a band inversion hosts an edge state. No matter how many band inversions occur, as long as it is odd in number, we have the same $Z_2$ invariant $\nu_0 = 1$, but the corresponding number of Dirac cones at the edge is equal to the number of band inversions. This is in contrast to the Chern insulator where the bulk topological invariant dictates the number of surface state.

## H. $Z_2$ calculation *without* inversion symmetry

Subsequently, Fu and Kane have generalized the $Z_2$ calculation for systems without inversion symmetry[97]:

$$\nu = \frac{1}{2\pi} \left[ \oint_{d\tau} A_n(\mathbf{k}) dk - \int_{\tau} F_n(\mathbf{k}) d^2k \right], \quad (17)$$

$\tau$ is half of the BZ one in which the Berry curvature is to be computed, while $d\tau$ is the boundary where the integral of the Berry connection is to be calculated (see Fig. 8). The difference between Eq. (17) and Eq. (5) is that here an additional surface integral over the Berry connection is present. This term appears in the process of gauge fixing as follows.

In systems with finite Chern number, the center of the cyclotron orbit or 'chiral orbit' poses an obstruction to smoothly affix a gauge to the wavefunction. Because, the phase of the wavefunction is supposed to acquire a discontinuity at the center of the orbit to commence finite phase winding or Chern number. If the wavefunction has a smooth gauge at all k-points, both **A** and **F** acquire the same gauge, yielding $\nu = 0$ by Stokes' theorem. For $Z_2$ TIs, although the Chern

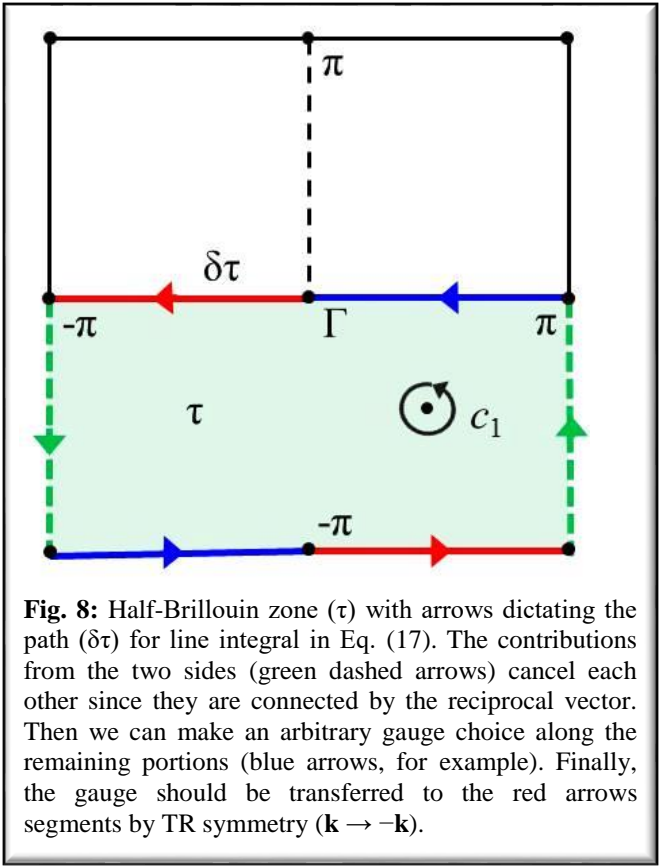

**Fig. 8:** Half-Brillouin zone ($\tau$) with arrows dictating the path ($\delta\tau$) for line integral in Eq. (17). The contributions from the two sides (green dashed arrows) cancel each other since they are connected by the reciprocal vector. Then we can make an arbitrary gauge choice along the remaining portions (blue arrows, for example). Finally, the gauge should be transferred to the red arrows segments by TR symmetry ($\mathbf{k} \to -\mathbf{k}$).

$$\nu_L = \frac{1}{2i\pi}\left[\sum_{\mathbf{k}\in\delta\tau} A_1(\mathbf{k}) - \sum_{\mathbf{k}\in\tau} F_1(\mathbf{k})\right]. \quad (18)$$

For systems with inversion symmetry, this formula cannot be used since $A$ and $F$ terms cancel each other.

The 3D generalization of this calculation follows similarly.[10] A 3D BZ has six inequivalent 2D planes at $k_i = 0, \pi$, where $i = x, y, z$. Using Eq. (18), one obtains six topological invariants on these six planes, namely $\nu^i_{0,\pi}$. Among which, the strong topological invariant condition implies that $\nu^x_0 \nu^x_\pi = \nu^y_0 \nu^y_\pi = \nu^z_0 \nu^z_\pi$. Therefore, as in Sec. IIF, we get four independent topological invariants defined by $\nu_0 = \nu^x_0 \nu^x_\pi$, $\nu_1 = \nu^x_\pi$, $\nu_2 = \nu^y_\pi$, $\nu_3 = \nu^z_\pi$.

### I. Axion angle as topological invariant

According to the axion electrodynamics,[101] electric and magnetic fields can couple linearly, giving rise to an additional term in the Maxwell's action $S_\theta = \left(\frac{\theta}{4\pi}\right)\left(\frac{\alpha}{2\pi}\right)\int d^3x\, dt\, \mathbf{E}.\mathbf{B}$, where $\alpha$ is the hyperfine constant, and $\theta$ is the coupling constant, called the axion field. Given that the Berry curvature $\mathbf{F}(\mathbf{k})$ acts like a momentum space magnetic field, a similar effect can be expected here in that $\mathbf{F}(\mathbf{k})$ can couple to the polarizibility of the charged particles. This is what is shown by the Chern-Simon theory,[14,24,102–104] which can in fact describe the IQH effect. In this case, the Maxwell's action becomes $S_\theta = \left(\frac{\theta}{32\pi^2}\right)\int d^3x\, dt\, \varepsilon^{\mu\sigma\tau\rho}F_{\mu\sigma}F_{\tau\rho}$. The axion angle can be computed from the Berry connection [14,24,104,105]:

$$\theta = \frac{1}{4\pi}\int d^3k\, \varepsilon^{\mu\sigma\tau}\, \text{Tr}\left[A_\mu \partial_\sigma A_\tau - i\frac{2}{3}A_\mu A_\sigma A_\tau\right], \quad (19)$$

number is zero, but a similar obstruction arises at the nodes of the Pfaffian where the wavefunction switches to its TR conjugate [see Sec. IIF]. This affirms that one cannot simultaneously obtain a node in the Pfaffian and a smooth gauge in the wavefunction.

However, Fu and Kane proposed[97] that if the contour is restricted within half of the BZ, see Fig. 8, and also we choose a gauge that is periodic, i.e., $|u(\mathbf{k})\rangle = |u(\mathbf{k}+\mathbf{G})\rangle$, in addition to the TR symmetry, the problem can be solved, see discussions in Refs. [10,98–100]. In this case, Eq. (17) can be solved by converting the integral into summation over a uniform discretized $\mathbf{k}$-mesh on the region $\tau$. Imposing the TR symmetry on the boundary $d\tau$, we can obtain a link matrix as $W^{mn}_i(\mathbf{k}) = \langle u_m(\mathbf{k})|u_n(\mathbf{k}+\delta\mathbf{k}_i)\rangle$, and the unimodular link variable $\Delta_i(\mathbf{k}) = \det W_i/|\det W_i|$, where $\delta\mathbf{k}_1$ ($\delta\mathbf{k}_2$) is the step of the mesh in the direction of the reciprocal vectors $\mathbf{G}_1$ ($\mathbf{G}_2$). Then we define the gauge potential as $A_i(\mathbf{k}) = \log \Delta_i(\mathbf{k})$, which gives $F(\mathbf{k}) = \log[\Delta_1(\mathbf{k})\Delta_2(\mathbf{k}+\delta\mathbf{k}_1)\Delta_1(\mathbf{k}+\delta\mathbf{k}_2)^{-1}\Delta_2(\mathbf{k})^{-1}]$. Then the $Z_2$ invariant can be calculated as

As in the case of IQH effect, $\theta$ becomes quantized to be 0 or $\pi$ (mod $2\pi$) if the system is TR invariant. For $Z_2$ TI, $\theta$ reflects the same topological invariant as $\theta = \nu\pi$. Therefore, $\theta = 0$ indicates a topologically trivial state, while $\theta = \pi$ signifies a non-trivial topology. Interestingly, while $\nu$ changes by integer values, $\theta$ changes continuously from $\pi$ to 0 as TR symmetry is lifted by spontaneously introducing magnetic moment. Again unlike the $Z_2$ invariant, axion field $\theta$ is an observable since it gives rise to a topological electromagnetic effect. Therefore, TI with small magnetic moment belongs to a new class, namely topological axion insulator. So far, this state is proposed in few materials,[25,105,106] but not realized yet.

As the magnetic moment is further increased, another class of TI, called quantum anomalous Hall (QAH) insulator, may arise if the system undergoes a similar band inversion. Here, a $Z_2$ classification is destroyed, and the system can in principle possess arbitrarily large value of Chern number. QAH state is proposed and realized in various engineered structures,[61,107–109] and La$X$ (X=Br, Cl, I) is the only family predicted so far as intrinsic QAH insulator.[62]

### J. Topological invariant for interacting fermions

For non-interacting systems, the topological invariant can be extracted from the Kubo formula for the Hall conductivity in 2D [Eq. (3)]. For interacting systems, one can follow the same strategy.[89,110] For such systems, a single-particle wavefunction cannot be defined and thus we start from a different Kubo formula for interacting systems as $\sigma_{xy} = \frac{e^2}{\hbar} \text{Im} \frac{\partial}{\partial \omega} K(\omega + i\delta)$. The current-current correlation kernel $K(\omega + i\delta)$ can be expressed in terms of the interacting Green's function $G(k, i\omega)$ as

$$K(i\omega) = -\frac{1}{\Omega_{\text{BZ}} \beta} \sum_{k, i\nu} \text{Tr}[v_x G(k, i\omega + i\nu) v_y G(k, i\omega)], \quad (20)$$

where $\Omega_{\text{BZ}}$ is the phase space volume, $\beta = 1/k_B T$, and the velocity vertices are $v_i(\mathbf{k}) = \partial H(\mathbf{k})/\partial k_i$. For (2+1)D systems, the vertex can be expressed within the Ward identity as $v_i(p) = \partial G^{-1}(p)/\partial p_i$, where $p = (\mathbf{k}, i\omega)$. Integrating over the (2+1)D phase space volume, we obtain a similar topological invariant (Chern number) in terms of generalized Green's function as[90]

$$\nu = \frac{\pi}{6} \int \frac{d^3 p}{(2\pi)^3} \text{Tr} \left[ \varepsilon^{\mu\rho\sigma} G \frac{\partial G^{-1}}{\partial p_\mu} G \frac{\partial G^{-1}}{\partial p_\rho} G \frac{\partial G^{-1}}{\partial p_\sigma} \right]. \quad (21)$$

Extension to higher dimensions follows the same procedure, in which the number of vertices is equal to the dimension of the system.[14,111] Clearly, Eq. (21) is also applicable to non-interacting Green's function $G_0(k, i\omega) = (i\omega - H(k))^{-1}$. In the multi-orbital systems, corresponding Green's function is a tensor. Electron-electron interaction or disorder effect can be incorporated within the Dyson, or the $T$ - matrix formalism, among others, giving a generalized formalism $G(k, i\omega)^{-1} = G_0(k, i\omega)^{-1} - \Sigma(k, i\omega)$, where $\Sigma$ is the self-energy correction. Dynamical mean-field theory (DMFT), and momentum-resolved density fluctuation (MRDF) theory[112–117] are two widely used methods to explore the dynamical correction effects. The latter method has an added advantage of incorporating the full momentum dependence of the correlation effects. For many systems such as transition metal oxides,[112,115] and di-chalcogenides,[114] intermetallics, lanthanum and actinide compounds,[113,116] the momentum dependence of the electron-correlation is significantly strong, which can lead to a characteristic change in the Berry curvature $\mathbf{F}(\mathbf{k})$. These features can be captured within the MRDF method.

### K. Topological invariants for superconductors

Superconductivity is a correlated phenomenon which arises due to the condensation of electron-electron pair (Cooper pair) in the low-energy spectrum. Within the mean-field theory, the corresponding Hamiltonian can be casted into a single particle (quasiparticle) Hamiltonian in which the superconductivity opens a band gap at the Fermi level. Therefore, although in the two electrons picture, the system is superconducting (SC), in the single electron effective model it represents an insulator (assuming the SC gap opens everywhere on the Fermi surface). Interestingly, fully gapped superconductor, and topological insulator share an analogous Hamiltonian, and thus many of the topological concepts also applies in the former case.[22,83,118–123] The single-band Hamiltonian for a superconductor can be expressed exactly by Eq. (9), with $d_3 = \varepsilon(\mathbf{k})$, $d_{1,2}$ are the real and imaginary parts of the SC gap, $\Delta(\mathbf{k})$. Furthermore, the owing to the criterion for the formation of 'chiral orbit' or 'chiral vortex', the SC gap must be a chiral pairing symmetry, which is often obtained in $p$-wave superconductors (for spinful superconductors, this condition can be relaxed if SOC is present[23]). The chiral $p$-wave superconductors have odd parity gap symmetry and breaks TR symmetry. In such a case, the topological invariant is obtained by the sum of the first Chern number ($\nu_n$) on each band weighted by the sign of the gap as[22,23,118,124]

$$\mathcal{N} = \frac{1}{2} \sum_n \nu_n \, \text{sgn}(\Delta_{n\mathbf{k}}). \quad (22)$$

For spinful case, TR invariant topological superconductor can be obtained if the pairings

$\langle \psi^\dagger_{k\uparrow} \psi^\dagger_{-k\uparrow} \rangle$ and $\langle \psi^\dagger_{k\downarrow} \psi^\dagger_{-k\downarrow} \rangle$ have opposite chirality, i.e., $\Delta_{\uparrow\uparrow} = p_x + ip_y$, and $\Delta_{\downarrow\downarrow} = p_x - ip_y$. Here again if the corresponding $4 \times 4$ Hamiltonian can be split into the block diagonals, as in the case of half-BHZ model for QSH insulator, we can apply the same Chern number calculation to evaluate the topological invariant. Due to the associated particle-hole symmetry, the zero energy boundary modes must be a Majorana mode,[83,119,123] which means, its eigenstate must be real.

**L. Adiabatic continuity**

Adiabatic continuation is a simple and powerful tool to identify a non-trivial TI with reference to another known TI, if both these systems are adiabatically connected. Here 'adiabatic connection' simply means that as one transforms a non-trivial TI 'A' into another material 'B' by *continuously* changing the atomic number of the constituent elements, the bulk band gap of the 'A' system does not close and reopen in this whole process, then they are adiabatically connected or belong to the same non-trivial TI class.

The band evolution between a non-trivial TI to a trivial TI is shown in Fig. 9. Suppose with a given tuning, such as chemical potential, or atomic number, or pressure, the band gap simply decreases, without any band inversion, as in going from Fig. 9(a) to 9(b). Then these two systems are adiabatically connected and belong to the same topological class. At the topological critical point, when the bulk band gap closes at the TR invariant point, it produces a bulk Dirac cone (also refereed as 3D Dirac cone), Fig. 9(c). Note that graphene is a non-trivial system lying at the topological critical point. Above the critical point, as the bands are inverted back, the system transforms into a trivial insulator. Such band evolution including the 3D Dirac cone formation is seen experimentally in BiTl(S$_{1-\delta}$Se$_\delta$)$_2$.[65]

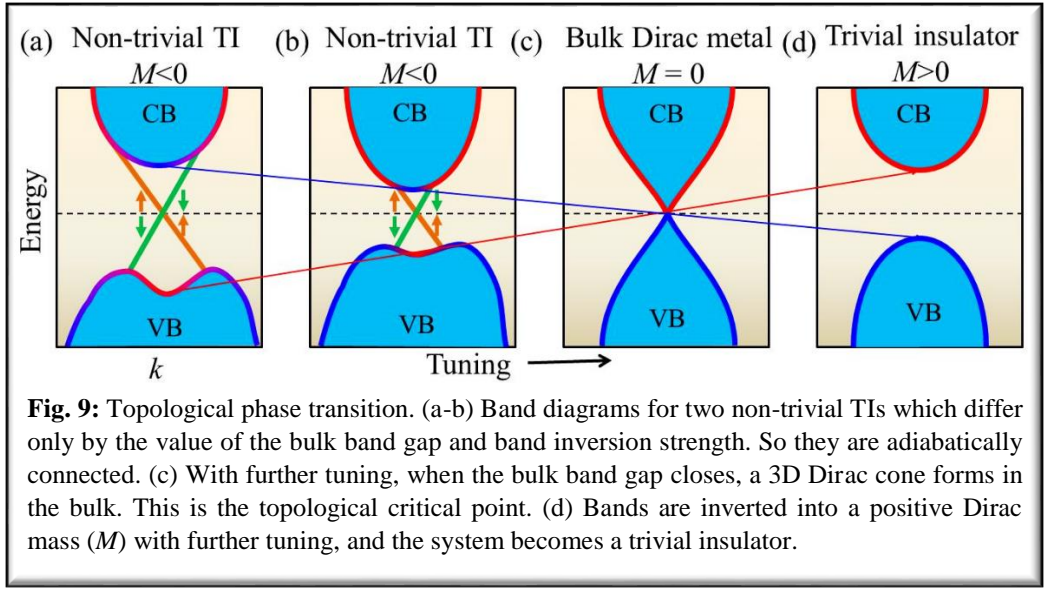

**Fig. 9:** Topological phase transition. (a-b) Band diagrams for two non-trivial TIs which differ only by the value of the bulk band gap and band inversion strength. So they are adiabatically connected. (c) With further tuning, when the bulk band gap closes, a 3D Dirac cone forms in the bulk. This is the topological critical point. (d) Bands are inverted into a positive Dirac mass ($M$) with further tuning, and the system becomes a trivial insulator.

The difficulty with this method is that one requires to study the band topology by continuously changing the atomic number, i.e. by doping, which is not an easy calculation within the first principles methods. Yet, the method has been successful in predicting a number of materials, especially those which does not have inversion symmetry, by starting with a nearby known TI which has inversion symmetry. Lin *et al.*, have shown[20,37] that one can test a large class of materials by adiabatically changing the atomic number of the constituent elements, which is done by alloying or doping. For example, we can start with a hypothetical system with three elements *MM'X*, with nuclear charge $Z_M = 3 - 0.5x + 0.5y$, $Z_{M'} = 47 + x$, and $Z_X = 51 - y$, respectively, where $x$ and $y$ are adjustable parameter. $x$ and $y$ are not necessarily integer, but the choice must maintain the charge neutrality. This mapping can start with $x = 0$ and $y = 0$, which corresponds to Li$_2$AgSb, and end with $x = 3$ and $y = 1$, which corresponds to the artificial compound He$_2$SnSn. Li$_2$AsSb is known to be a non-trivial TI. With increasing $x$, and $y$, we obtain Li$_2$AgBi, Li$_2$AuBi, and Li$_2$CdSn which are non-trivial topological metals. On the other hand, the end element Li$_2$CuSb is a trivial band insulator in which the bulk band gap has reopened above the critical point.

**M. Bulk-boundary correspondence and surface states**

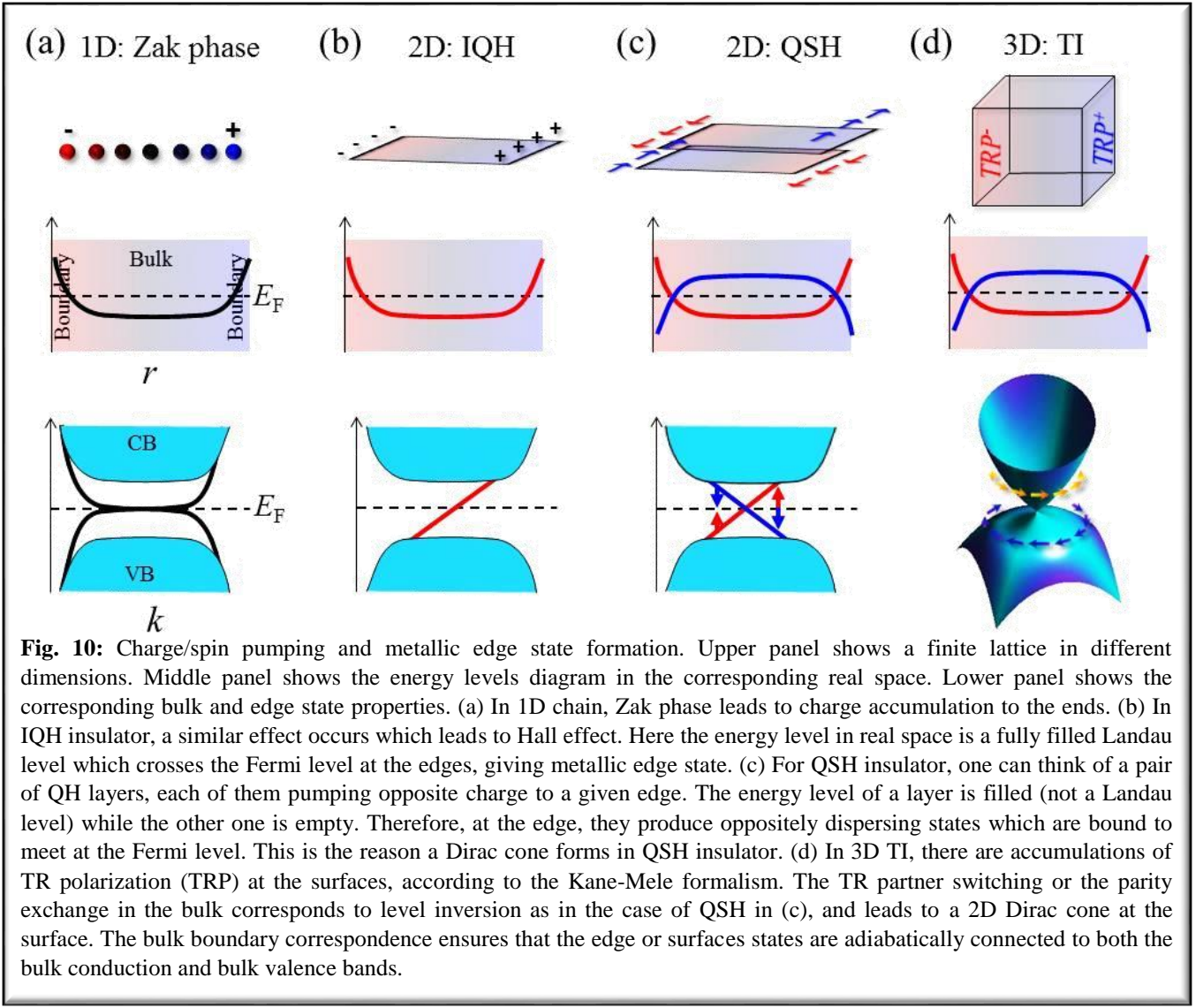

**Fig. 10:** Charge/spin pumping and metallic edge state formation. Upper panel shows a finite lattice in different dimensions. Middle panel shows the energy levels diagram in the corresponding real space. Lower panel shows the corresponding bulk and edge state properties. (a) In 1D chain, Zak phase leads to charge accumulation to the ends. (b) In IQH insulator, a similar effect occurs which leads to Hall effect. Here the energy level in real space is a fully filled Landau level which crosses the Fermi level at the edges, giving metallic edge state. (c) For QSH insulator, one can think of a pair of QH layers, each of them pumping opposite charge to a given edge. The energy level of a layer is filled (not a Landau level) while the other one is empty. Therefore, at the edge, they produce oppositely dispersing states which are bound to meet at the Fermi level. This is the reason a Dirac cone forms in QSH insulator. (d) In 3D TI, there are accumulations of TR polarization (TRP) at the surfaces, according to the Kane-Mele formalism. The TR partner switching or the parity exchange in the bulk corresponds to level inversion as in the case of QSH in (c), and leads to a 2D Dirac cone at the surface. The bulk boundary correspondence ensures that the edge or surfaces states are adiabatically connected to both the bulk conduction and bulk valence bands.

Spontaneous (continuous) symmetry breaking leads to gapless Goldstone mode in the corresponding excitation spectrum. For example, the spin rotational symmetry breaking in quantum magnets leads to gapless magnons in the spin excitation spectrum, or the translational symmetry breaking in the formation of a lattice renders gapless phonon mode (acoustic modes). Similarly, as the cyclotron orbits or 'chiral orbits' become periodically arranged in a lattice (creating a Gaussian curvature), breaking the translational symmetry, which is associated with the emergence of non-trivial bulk topology, it manifests into gapless edge states at the boundary. Although a rigorous calculation to validate this premise is yet not explored, however, the application of Goldstone theory for the realization of bulk-boundary correspondence can be intriguing. For example, electromagnetic response of TIs stipulates two dynamical axion modes, one of them is gapless Goldstone-like mode, and another is Higgs-like gapped mode, as shown in earlier calculation.[104]

According to the bulk-boundary correspondence of TI, the bulk topological invariant dictates the number and characteristics of edge states at the boundary. For the case of IQH effect, the Chern number $N$ prescribes $N$ chiral edge states. For the spin or mirror Chern numbers, edge states form in pair; for example, $N$ spin Chern number has $2N$ counter-propagating chiral edge states. The same principle also applies to $Z_2$ topological invariant with some modifications. Kane-Mele proposed that for TR invariant $Z_2$ TI, the TR partners are accumulated at different sides leading to the 'TR polarization'. In the presence of additional inversion symmetry, TR polarization is equivalent to 'parity polarization' in which different parity states are pumped to different sides. This also implies that the edge or surface state must contain both TR partners, which means that the edge states must come in pair. However both states must also be degenerate at the TR invariant **k**-points, forming a Dirac cone, which is guaranteed by the $Z_2$ invariant in the bulk.

With the application of perpendicular magnetic field, electrons and hole are pushed towards opposite side of a plane, leading to Hall effect. The theory of bulk-boundary correspondence for TIs can be developed similarly, by studying the effect of the momentum space 'magnetic field' or Berry curvature in the boundary. This is precisely what is done by Zak, by calculating the Berry phase in a finite 1D chain to obtain the so-called Zak phase or end state at the boundary. Zak[95] assumed a bulk periodic boundary condition in the momentum space which yields [from Eq. (5)],

$$\gamma_n = i\frac{2\pi}{a}\int_{-\pi/a}^{\pi/a} dk \int_0^a u_{nk}^*(x)\frac{\partial u_{nk}(x)}{\partial k}dx, \quad (23)$$

where $u_{nk}(x)$ is the periodic part of the Bloch state, and $a$ is the lattice constant. By using the corresponding Wannier function $a_n(x)$, he arrived at a spatial dependence of the Berry phase as

$$\gamma_n = \frac{2\pi}{a}\int_{-\infty}^{\infty} x|a_n(x)|^2\, dx. \quad (24)$$

We notice that the above equation is identical to that of the electric polarizibility when multiplied with electric charge. As the inversion symmetry is imposed, $\gamma_n$ takes the values of either 0 or $a/2$. At both ends, the integral range survives either $-\infty$ to 0, or 0 to $\infty$. Since the integrand is odd function of $x$, the integral obtains $\pm\gamma_n$ at the $\pm x$ ends, respectively. By multiplying elementary charge $e$, we see that the polarization at the two ends have opposite sings, implying that the electrons and holes are separated into different ends of the 1D chain. This is referred as charge pumping in 1D TI [see Fig. 8(a)]. In the 2D IQH effect, similarly, electron and hole states are accumulated at the two edges, see Fig. 8(b). Since a single edge state has to smoothly connect both the electron (filled band) and hole (empty band) state, it has to pass through the neutral or zero energy mode. Therefore, the zero energy edge state is guaranteed by the bulk topology, and cannot be destroyed by any weak perturbation or disorder, as long as bulk topology remains intact.

For the case of QSH insulator, opposite spin Chern numbers dictate opposite charge pumping to a given edge. For example, if the spin-up state drives electrons toward the +$x$ direction, then spin-down state will pump holes to the same side. This cancels the net charge in each edge but allows a net spin accumulation, see Fig. 10(c). Therefore, here always a pair of counter dispersive and spin-polarized edge states arises which meet at the TR invariant momentum, $k_s^*$, owing to Kramers' degeneracy. Let us assume an edge state along the $k_y$ direction. Since any band dispersion can be expanded via Taylor series around the $k_s^*$ as: $\varepsilon(q_y) = \varepsilon_s + q_y v_s^*$, with $q_y = k_y - k_{ys}^* \ll 1$, and $v_s^*$ is the band velocity. Since $\varepsilon_s$ is a constant term, we can neglect it. TR symmetry demands $\Theta|u_\uparrow(q_y)\rangle = |u_\downarrow(-q_y)\rangle$, implying that if we set $\varepsilon_\uparrow = q_y v_s^*$, the band for the opposite spin must be $\varepsilon_\downarrow = -q_y v_s^*$, or vice versa. For an electron below the Fermi level, its spin is interchanged as it traverses to the other side of $k_s^*$, or as $q_y \to -q_y$. Equal and opposite effect simultaneously occurs for the hole state. Therefore, if we express our edge Hamiltonian in the spinor $(\psi_\uparrow, \psi_\downarrow)$, the minimal, low-energy Hamiltonian for an 1D edge state with eigenvalues $\pm q_y v_s^*$ is[23]

$$H_{\text{edge}} = v_s^* q_y \sigma_z. \quad (25)$$

The corresponding edge state is schematically shown in Fig. 10(c) (lower panel).

We can understand the emergence of 2D surface state in 3D TI in a similar way, with the idea of 'TR polarization'.[97] Let us start with a stack of 2D QSH insulators placed along the $x$-direction, each of which containing topological edge states along the $y$-direction. Now as we turn this into a 'strong' 3D TI, in which the spin pumping (for TR polarization in general) must occurs along both the $y$-, and $x$-directions. This constraints the surface Hamiltonian to have 2D spin-polarization, and can therefore be written in the same spinor $(\psi_\uparrow, \psi_\downarrow)$ as [see Fig. 10(d)]

$$H_{\text{surf}} = v_s^* \mathbf{q}\cdot\boldsymbol{\sigma}. \quad (26)$$

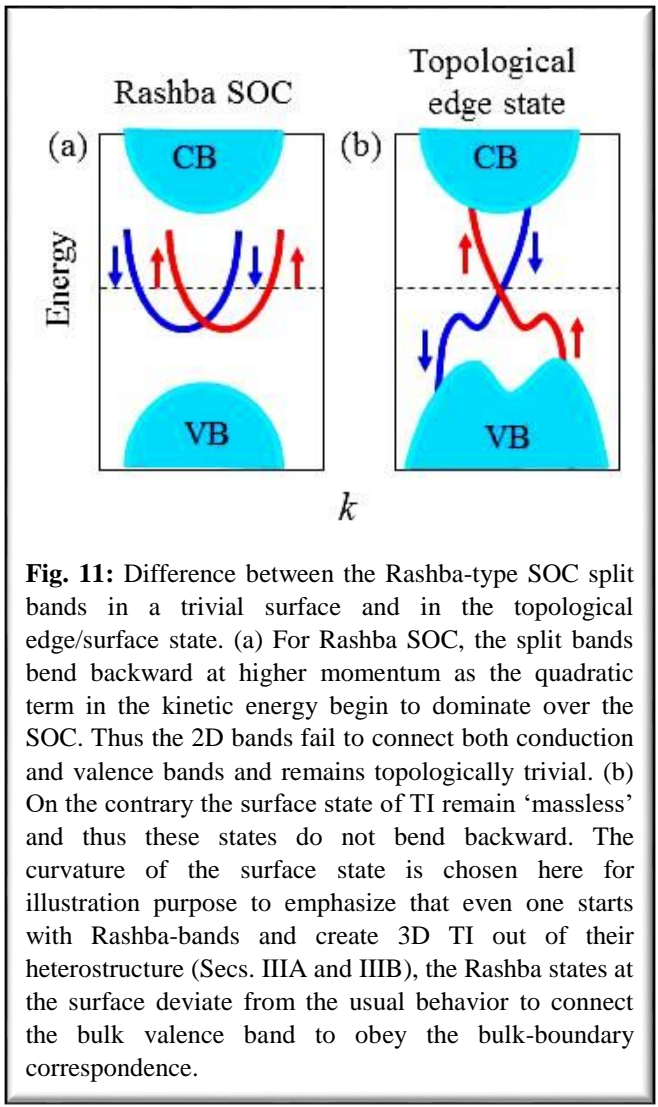

**Fig. 11:** Difference between the Rashba-type SOC split bands in a trivial surface and in the topological edge/surface state. (a) For Rashba SOC, the split bands bend backward at higher momentum as the quadratic term in the kinetic energy begin to dominate over the SOC. Thus the 2D bands fail to connect both conduction and valence bands and remains topologically trivial. (b) On the contrary the surface state of TI remain 'massless' and thus these states do not bend backward. The curvature of the surface state is chosen here for illustration purpose to emphasize that even one starts with Rashba-bands and create 3D TI out of their heterostructure (Secs. IIIA and IIIB), the Rashba states at the surface deviate from the usual behavior to connect the bulk valence band to obey the bulk-boundary correspondence.

Note that this surface Hamiltonian is very analogous to the Rashba-type SOC $h_R = k^2/2m\, I_{2\times 2} \pm \alpha_R(\mathbf{q}\times\boldsymbol{\sigma})$, where $\alpha_R$ is the SOC strength and the other symbols have the usual meanings. Indeed, owing to the loss of inversion symmetry at the surface, the surface states actually originate from the Rashba-type SOC, with a crucial difference. In the topological surface state, the kinetic energy of each spin ($k^2/2m$) is much smaller compared to the SOC strength, $k^2/2m \ll \alpha_R$. This is a manifestation of the bulk-boundary correspondence of the TI. For the Rashba-band, in Fig. 11(a), each spin band bends backward away from the TR momentum as the kinetic energy term dominates over SOC term at higher momenta. The Rashba-bands are not required to connect to the bulk states at all. On the other hand, topological surface states are required to adiabatically connect to both the bulk conduction and valence bands, and thus they cannot bend backward and rather disperse monotonically across the bulk insulating gap. This is illustrated in Fig. 11(b). It is obvious that all edge/surface states are not topological surface state. More importantly, even in a TI, all surface states are not necessarily topological surface state as well.[125] A topological surface state must have a Dirac cone at the TR point where band inversion occurred in the bulk state and it also has to connect both the bulk valence and conduction states. There might be situation where an apparently isolated topological state may arise which is not connecting any other state.[126] Here however, if one studies the eigenvector or weight of the band, it may appear that the weight has changed from surface to bulk as one moves away from the Dirac node.

## III. Engineering topological insulators

The above introduction highlights some key ingredients universally relating different TI families. The foremost ingredient is that electrons must contain a chirality, i.e., complex hopping term, which is obtained either via applied magnetic field (Pierels phase), or in bipartite lattice (such as SSH,[15] or graphene lattice,[94]), or via SOC, or in orbital selective lattice,[16,17] or even by artificial gauge fields as often done in optical lattice systems.[96,127] These chiral states must also form cyclotron or 'chiral orbits', with or without magnetic field. The 'chiral orbits' naturally produces a k-space 'magnetic fields', or Berry curvature. As these 'chiral orbits' are arranged in a 2D lattice, the system can be represented by a Gaussian curvature (torus as in the case of IQH system), each threading counter propagating integer multiple of quantum flux of the Berry curvature. In Sec. IIC, we also discussed that the 'chiral orbits' does not have to be formed in a physical parameter space, such as real or momentum space, but can also be formed in any generalized parameter space as long as they obey periodic boundary conditions. For such case, the Hall conductance can be precisely calculated. Therefore, for the cases where the two chiral states can be fully separated in the Hamiltonian (as in the case of half-BHZ model,[9] for example), each state can be assigned with spin or mirror Chern numbers, as applicable, with their net value vanishes, but the difference yields finite value. For the generalization to evaluating $Z_2$ invariant, odd pair of Pfaffian nodes is analogous to the center of odd pairs of 'chiral orbits' in the momentum space. Here one requires that the electron's wavefunction must change into its TR conjugate odd number of times as it traverses through half of the BZ[4,97]. These ingredients are used in the following methods as the targets to engineer TIs.

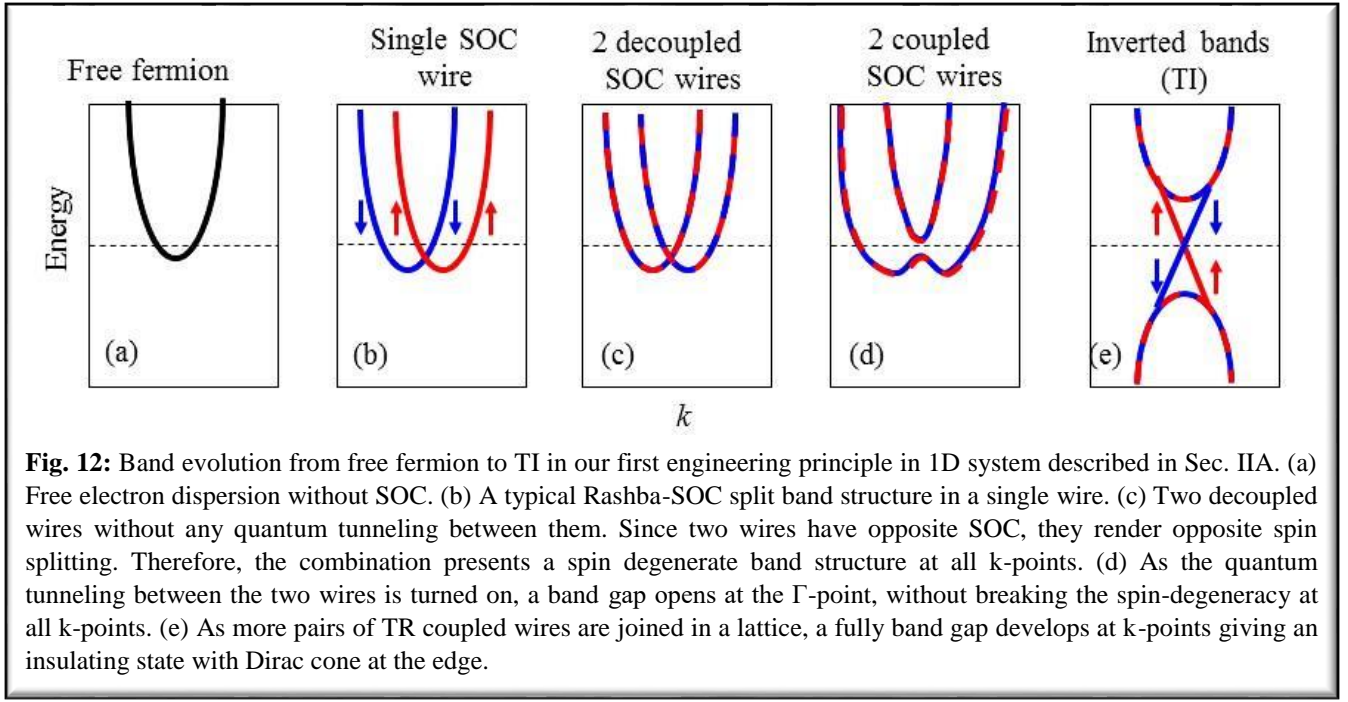

**Fig. 12:** Band evolution from free fermion to TI in our first engineering principle in 1D system described in Sec. IIA. (a) Free electron dispersion without SOC. (b) A typical Rashba-SOC split band structure in a single wire. (c) Two decoupled wires without any quantum tunneling between them. Since two wires have opposite SOC, they render opposite spin splitting. Therefore, the combination presents a spin degenerate band structure at all k-points. (d) As the quantum tunneling between the two wires is turned on, a band gap opens at the Γ-point, without breaking the spin-degeneracy at all k-points. (e) As more pairs of TR coupled wires are joined in a lattice, a fully band gap develops at k-points giving an insulating state with Dirac cone at the edge.

The idea would be to take a bottom-up approach to assemble TIs in desired dimensions. Here we start with an atomic chain or layer with SOC, and invert the SOC in its adjacent chain or layer so that it acts as the TR conjugate partner to the former one. Therefore, the electron switches its TR partner in hopping from one chain/layer to another. The Pfaffian $P(\mathbf{k})$ thereby acquires a node in between them, and the $Z_2$ invariant becomes 1. This setup can be achieved by aligning the direction of SOC in each chain or layer by manipulating lasers (in optical lattice) or by reversing electron field directions (for Rashba-SOC), or via interaction. As an insulating state occurs, the system is guaranteed to behave as a $Z_2$ TI.

In systems with inversion symmetry, $Z_2$ invariant can be evaluated by the odd number of band inversion between the two chiral states at the TR invariant momenta.[7] This phenomena can be thought of as the 'chiral orbits' are split along the energy axis and/or in the momentum space with negative Dirac mass. At the end of this section, we will introduce another new concept for chiral band inversion in the real-space, due to interaction effect. We attribute the corresponding emergent topological phase as quantum spin-Hall density wave (QSHDW) insulator.[30] Our research group has a major thrust in this research direction, among which we give below four representative examples.

**A. Engineering topological 'chiral orbits' in 2D**

In the first example, we start with a 1D chain of atoms with 1D SOC. Such state arises in quantum wires,[128] optical lattices,[79,129,130] as well as in bulk systems when Rashba- and Dresselhaus-type of SOCs have equal strength. We denote such as wire by 'A' SO wire. The corresponding Hamiltonian in the continuum limit is[79,128–130]

$$H_A = \frac{k^2}{2m}I_{2\times 2} + i\alpha_R k_x \sigma_x, \qquad (27)$$

where the first term gives the kinetic energy, and the second term gives the SOC with strength $\alpha_R$. (In optical lattice, in the process of creating SOC with lasers, a Zeeman-like terms also arise which has the form of $\Omega\sigma_z$, where $\Omega$ depends on the laser strength[129]). The corresponding spin split state is shown in Fig. 12(b). So far researchers are only able to generate 1D SOC in optical lattice which is inadequate to create 'chiral orbit' in 2D plane and thus poses a serious setback to obtain 2D or 3D TIs here. We propose to use the second SOC wire (called 'B' wire) with opposite SOC such that its Hamiltonian can be written as $H_B(k) = H_A^*(-k)$. Without quantum tunneling between the two wires, the spin-polarization of bands for two wires are reversed, and the Γ-point has now four-fold degeneracy (see Fig. 12(c)). Therefore, bands at the Γ-point can now be gapped without breaking the TR symmetry (spin degeneracy is still present), by switching on the inter-wire quantum tunneling, as shown in Fig. 12(d). Note that in this setup we have obtained chirality along the $k_x$ direction through SOC, but the same thing is still lacking for hopping between the wires. This can be obtained by allowing staggered inter-wire hopping, in which the distance between 'A' and 'B' wires and that between 'B' and next 'A' wires ('A' wire lying on opposite sides of 'B' wire) can be made different.

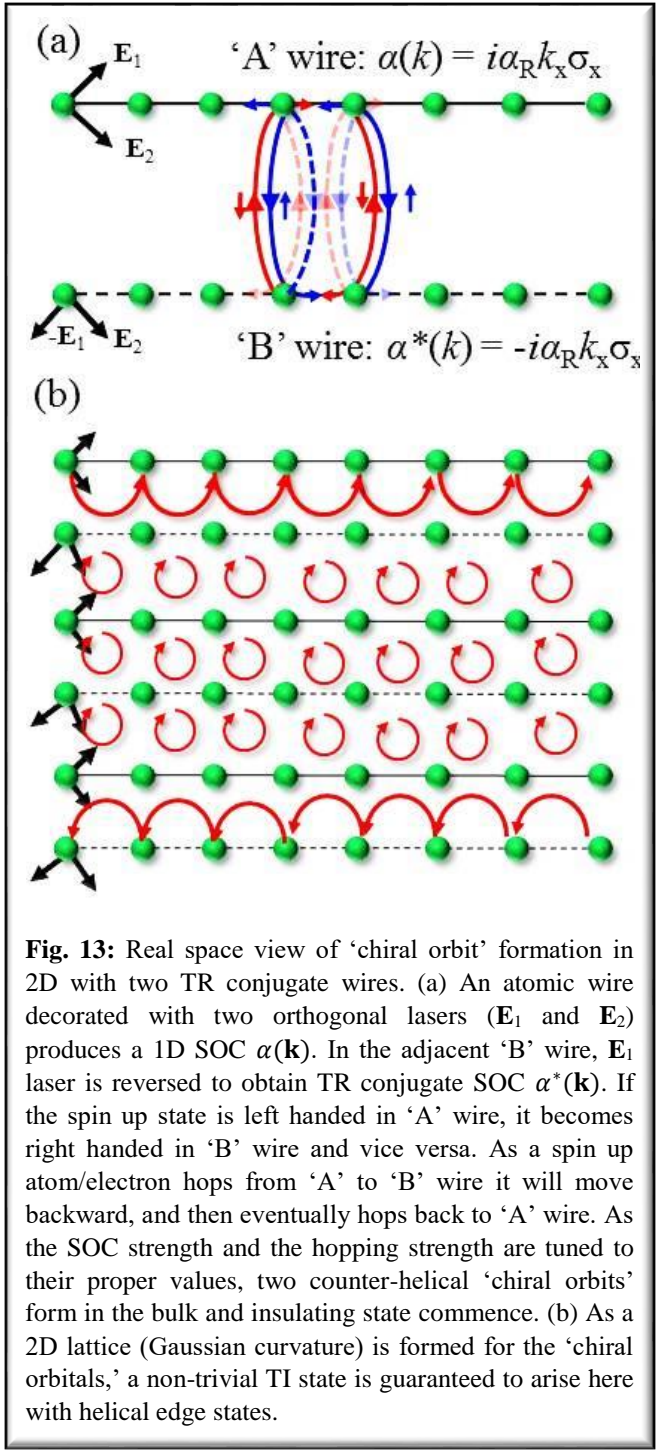

**Fig. 13:** Real space view of 'chiral orbit' formation in 2D with two TR conjugate wires. (a) An atomic wire decorated with two orthogonal lasers ($E_1$ and $E_2$) produces a 1D SOC $\alpha(\mathbf{k})$. In the adjacent 'B' wire, $E_1$ laser is reversed to obtain TR conjugate SOC $\alpha^*(\mathbf{k})$. If the spin up state is left handed in 'A' wire, it becomes right handed in 'B' wire and vice versa. As a spin up atom/electron hops from 'A' to 'B' wire it will move backward, and then eventually hops back to 'A' wire. As the SOC strength and the hopping strength are tuned to their proper values, two counter-helical 'chiral orbits' form in the bulk and insulating state commence. (b) As a 2D lattice (Gaussian curvature) is formed for the 'chiral orbitals,' a non-trivial TI state is guaranteed to arise here with helical edge states.

For this setup, the formation of localized 'chiral orbit' can be understood from the corresponding real-space view of the hoppings, see Fig. 13. Let us assume that a spin-up electron in the 'A' wire is right-moving, and that in the 'B' wire becomes left-moving, due to opposite SOC. Therefore, when a spin-up electron hops from 'A' wire to 'B' wire, its motion is reversed and as the electron hops back to the 'A' wire, it forms a 'chiral orbit' (reverse direction for the spin-down electron) as shown in Fig. 13(a). As the hopping amplitude is increased, these 'chiral orbits' become localized in the bulk, leading to a band insulator, as shown in Fig. 13(b). Due to the translational symmetry in the lattice, the TKNN invariants could have been assigned to each chiral state, with equal and opposite Chern numbers. However, due to SOC, the two chiral states switch TR partner across Γ-point. Therefore, we can calculate the 'TR polarization' as discussed in Sec. IIF. Finally, with the calculation of Pfaffian, we show that the present engineered device becomes a non-trivial TI, with spin polarized edge state as shown in Fig. 12(e). We find that the resulting Pfaffian[4,97] takes the form of $P(\mathbf{k}) = 1 + e^{-i2k_y}$, which therefore depends on the Bloch phase associated with the inter-wire hopping $e^{ik_y}$. The Pfaffian has one pair of nodes inside the first BZ at $k_y^* = \pm\pi/2$. Therefore, irrespective of the parameters, the present setup is guaranteed to produce a non-trivial TI.

**B. Engineering 3D topological insulators with Rashba bilayers**

Next we discuss our method of engineering 3D TIs by stacking 2D layers in a heterostructure, in which alternating layers have opposite SOCs.[5] Here we start with a 2D electron gas (2DEG) with Rashba-type SOC $\alpha(\mathbf{k})$. The approach is based on growing bilayer of Rashba-type 2DEG with opposite SOC on adjacent planes of bilayers. We find that in the stack of bilayers grown along (001)-direction, a topological phase transition occurs above a critical number of Rashba-bilayers, with the formation of a single spin-polarized Dirac cone at the Γ-point.

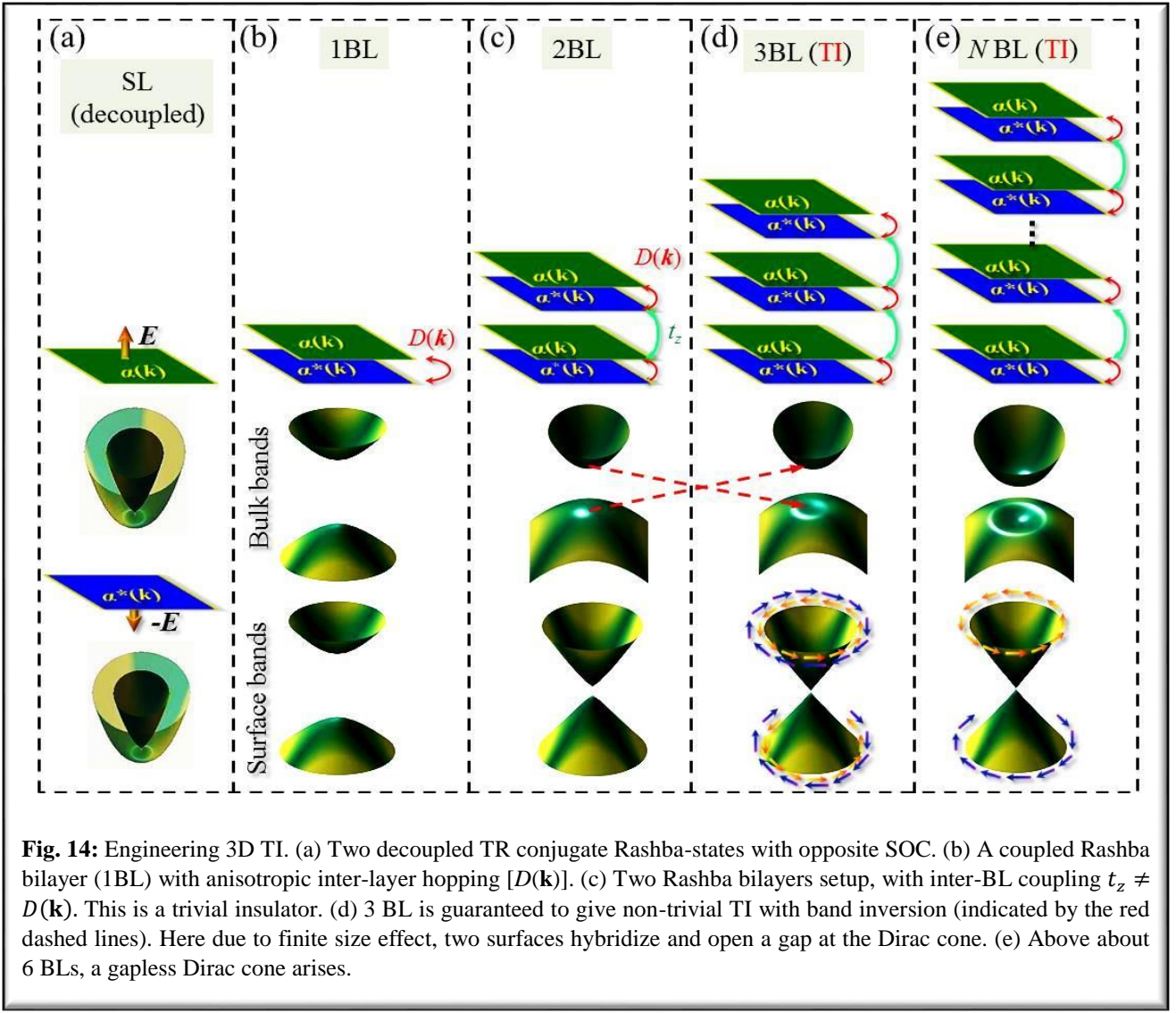

**Fig. 14:** Engineering 3D TI. (a) Two decoupled TR conjugate Rashba-states with opposite SOC. (b) A coupled Rashba bilayer (1BL) with anisotropic inter-layer hopping [$D(\mathbf{k})$]. (c) Two Rashba bilayers setup, with inter-BL coupling $t_z \neq D(\mathbf{k})$. This is a trivial insulator. (d) 3 BL is guaranteed to give non-trivial TI with band inversion (indicated by the red dashed lines). Here due to finite size effect, two surfaces hybridize and open a gap at the Dirac cone. (e) Above about 6 BLs, a gapless Dirac cone arises.

The band progression from trivial insulator to non-trivial TI is demonstrated in Fig. 14. The building block is a Rashba bilayer with opposite Rashba SOC in each layer, denoted by $\alpha(\mathbf{k})$ and $\alpha^*(\mathbf{k})$. We tune the interlayer distance such that an anisotropic quantum tunneling, $D(\mathbf{k})$, couples them, as illustrated in Fig. 14(b). Here again, without breaking the TR symmetry, the Rashba-bilayer opens an insulating gap at the Γ-points, which is determined by $D(0)$. Then we add another Rashba-bilayer on top of the previous one with an inter-bilayer electron hopping, $t_z$, which is required to be different from $D(0)$ to eliminate the degeneracy related to the number of bilayers. For two Rashba bilayers, the bulk bands in the two interior single layers have reduced band gap, and a massive 'preformed' Dirac like surface state appears, see Fig. 14(c). However, our parity analysis reveals that this setup still remains trivial insulator without any band inversion. However, as we add one more bilayer, a magic topological phase transition occurs, see Fig. 14(d). We see in the band structure that there exists a bulk band inversion between the valence and conduction bands at the Γ-point [indicated by arrows between Figs. 14(c), and 14(d)]. The band inversion can be easily visualized, confirmed by the parity analysis, from the change of curvature of the valence band near the Γ-point (or a 'dent' band structure). While the valence band is hole-like at all other momenta, it changes the topology to become electron-like at the Γ-point. Despite non-trivial band topology, the Dirac cone in the surface state obtains a tiny band gap due to finite size effect. As we grow a large size heterostructure with more Rashba bilayers, we see that band inversion strength in the bulk increases and the gap in the surface state gradually vanishes, see Fig. 14(e).

The low-energy effective model for a single Rashba-bilayer can be expressed as

$$H_{\rm BL}({\bf k}) = \begin{pmatrix} \varepsilon({\bf k}) & \alpha({\bf k}) & 0 & 0 \\ \alpha^*({\bf k}) & \varepsilon({\bf k}) & D({\bf k}) & 0 \\ 0 & D({\bf k}) & \varepsilon({\bf k}) & \alpha^*({\bf k}) \\ 0 & 0 & \alpha({\bf k}) & \varepsilon({\bf k}) \end{pmatrix},$$
(28)

with $\varepsilon({\bf k}) = k^2/2m$, the intra-bilayer hopping $D({\bf k}) = (D_0 + D_1 k^2)$. Each such bilayer is now stacked along the z-direction, which are connected by the nearest-neighbor inter-bilayer hopping $T = t_z I_{2\times 2}$. Interestingly, the resulting insulating bulk gap depends on the two tunneling terms $D_0$ and $t_z$, which are readily tunable. The surface state is determined by $H_{\rm surf} = \alpha_R(\sigma_x k_y - \sigma_y k_x)$, with its surface Dirac fermion velocity fully controlled by the Rashba SOC strength. For values of the Rashba-coupling constant as large as 3.8 eVA, achieved to date in bulk BiTeI,[131] we get $v = 5.8 \times 10^5$ ms$^{-1}$, which is much larger than the highest speed achieved so far in 3D TIs as $v \sim 3 \times 10^5$ ms$^{-1}$.[5]

Such Rashba-bilayer can be easily manufactured by creating a potential gradient between two 2DEGs with the help of gating, or by inserting oppositely polarized ferroelectric substrate between them, among others.[5] This idea is applied in the GaAs/Ge/GaAs heterostructure with opposite semiconductor interfaces.[132] The giant electric field generated by charge accumulation at the interfaces creates a Rashba-bilayer on both sides on Ge layer, and that allows a band inversion with an insulalting gap of 15 meV or larger. Another example is the synthesization of bulk strong TI by stacking 2D weak TIs[133]: Bi$_{14}$Rh$_3$I$_9$. Its Bi–Rh sheets are graphene analogues, but with a honeycomb net composed of RhBi$_8$ cubes rather than carbon atoms. The strong bismuth-related SOC renders each graphene-like layer be a TI with a 2,400 K bandgap. The non-centrosymmetric TlTeCl is an experimentally realized 3D TI[44] in which alternating layers of Bi and Te obtains opposite charge polarization, and thus opposite SOC. Therefore, our design principle is not only applicable in engineered heterostructure, but is also at play in bulk single crystals.

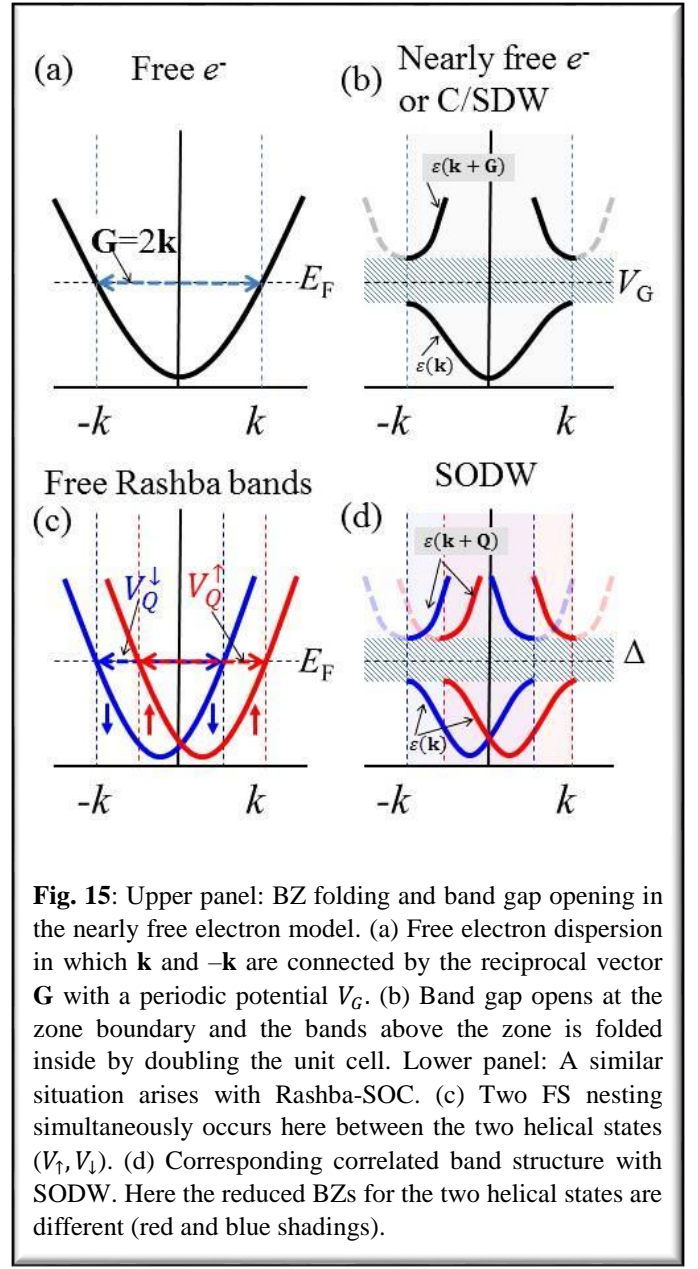

**Fig. 15**: Upper panel: BZ folding and band gap opening in the nearly free electron model. (a) Free electron dispersion in which **k** and –**k** are connected by the reciprocal vector **G** with a periodic potential $V_G$. (b) Band gap opens at the zone boundary and the bands above the zone is folded inside by doubling the unit cell. Lower panel: A similar situation arises with Rashba-SOC. (c) Two FS nesting simultaneously occurs here between the two helical states ($V_\uparrow$, $V_\downarrow$). (d) Corresponding correlated band structure with SODW. Here the reduced BZs for the two helical states are different (red and blue shadings).

## C. Spatial modulation of 'chiral orbits' and quantum spin Hall density wave insulator

Next we discuss an interesting situation where the energy level inversion occurs in real space between sublattices due to electron-electron interaction. To understand this phenomenon, we revisit the nearly free-electron model from elementary condensed matter course. We know that when a weak periodic potential $V_G$, where $G$ is the reciprocal wavevector, is applied to free electron gas, band gap opens at those $k$ - points which obey $2\mathbf{k} = \mathbf{G}$ relation, see Fig. 15(a-b). In this case, those parts of the bands, $\varepsilon(\mathbf{k})$, which lie above $2k > G$ are folded back into the reduced BZ by explicitly including it in the Hamiltonian as $\varepsilon(\mathbf{k}+G)$. The resulting system breaks translational symmetry, and thus opens a band gap at the zone boundary by $V_G$. A similar translational symmetry can occur due to electron-electron or electron-phonon interactions generated potential at a preferential wavevector $V_Q$, where $Q$ is called the Fermi surface nesting vector, defined as $2\mathbf{k}_F = \mathbf{Q}$, where $\mathbf{k}_F$ is the Fermi momenta. As a result, charge or spin density of the electrons at each lattice site becomes modulated and obtain a new periodicity which is different from the original periodic lattice. Similarly, at the edge of the reduced BZ [at $\mathbf{k} = \mathbf{Q}/2$], a quasipartcle gap opens. Here the gap is defined by the Landau-like order parameter, and the corresponding states are called charge/spin-density wave orders.

An analogous, but more exotic, situation arises when the Fermi surface nesting occurs between the two SOC split bands. As shown in Fig. 15(c-d), in such cases, the nesting simultaneously occurs for both spin states, giving rise to the spin-resolved potentials $V_Q^{\uparrow\downarrow}$ (unless TR symmetry is broken, the absolute values of the two potential can be equal). For such a case, there will be two spin-density waves, with each spin density having opposite spin-polarization at a given site, due to SOC. In such a case, a distinct order parameter, namely spin-orbit density wave (SODW) arises, which breaks translational symmetry, but in most cases it preserves TR symmetry. A detailed discussion of this order parameter can be found in Refs. [31,32,59,128,134]. Here we discuss how such a state can naturally give a $Z_2$ topological order parameter.[30]

We illustrate this case for a Rashba-type SOC. Inside the reduced zone, the non-interacting band, and the Rashba SOC are $\varepsilon(\mathbf{k})$, and $\alpha(\mathbf{k})$, and their folded counterparts are $\varepsilon(\mathbf{k} + \mathbf{Q})$, and $\alpha(\mathbf{k} + \mathbf{Q})$. An interesting case arises when the nesting vector is exactly $\mathbf{Q} = (\pi, 0)/(0, \pi)$. In this case, the folded Rashba SOC changes to the complex conjugate of the main SOC as $\alpha(\mathbf{k} + \mathbf{Q}) = \alpha^*(\mathbf{k})$. [This becomes obvious if we use the lattice form of the Rashba SOC where it takes the form of $\alpha(\mathbf{k}) = \alpha_R(\sin k_y \sigma_x + \sin k_x \sigma_y)$, which changes to $\alpha^*(\mathbf{k})$ as $\mathbf{k} \to \mathbf{k}+\mathbf{Q}$.]

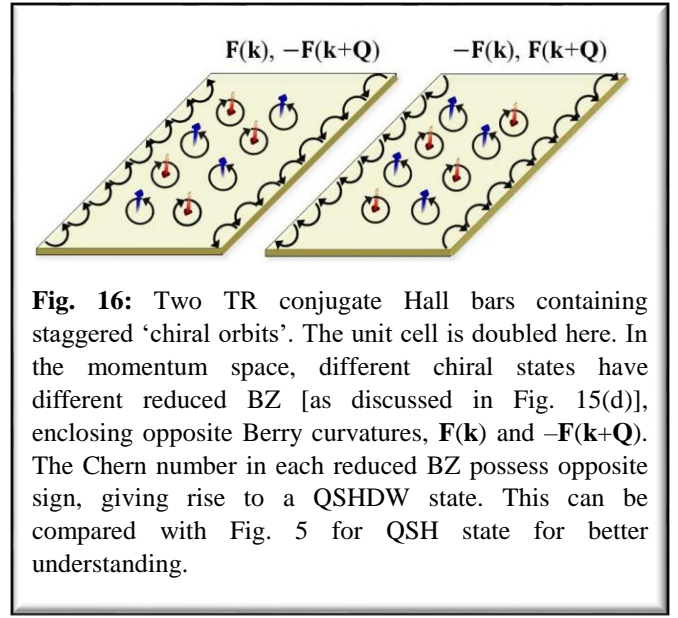

**Fig. 16:** Two TR conjugate Hall bars containing staggered 'chiral orbits'. The unit cell is doubled here. In the momentum space, different chiral states have different reduced BZ [as discussed in Fig. 15(d)], enclosing opposite Berry curvatures, **F(k)** and –**F(k+Q)**. The Chern number in each reduced BZ possess opposite sign, giving rise to a QSHDW state. This can be compared with Fig. 5 for QSH state for better understanding.

This gives rise to a situation which is analogous to the Hamiltonian for TI written in Eq. (28). To see that we can express a mean-field Hamiltonian in the basis of $\Psi_k = (\psi_{k\uparrow}, \psi_{k\downarrow}, \psi_{k+Q\uparrow}, \psi_{k+Q\downarrow})$ as

$$H = \begin{pmatrix} \varepsilon(\mathbf{k}) & \alpha(\mathbf{k}) & 0 & \Delta(\mathbf{k}) \\ \alpha^*(\mathbf{k}) & \varepsilon(\mathbf{k}) & \Delta^*(\mathbf{k}) & 0 \\ 0 & \Delta(\mathbf{k}) & \varepsilon(\mathbf{k}+\mathbf{Q}) & \alpha^*(\mathbf{k}) \\ \Delta^*(\mathbf{k}) & 0 & \alpha(\mathbf{k}) & \varepsilon(\mathbf{k}+\mathbf{Q}) \end{pmatrix}.$$
(29)

By comparing Eq. (28) with Eq. (29), we see that the single-electron tunneling term $D(\mathbf{k})$ in Eq. (29), is replaced here by the interacting SODW gap $\Delta(\mathbf{k}) = V\langle \psi_{k\uparrow}^\dagger \psi_{k+Q\downarrow}\rangle$, with $V$ being the interaction strength. The above Hamiltonian can also be expressed in terms of the Dirac matrices, in which the coefficient for the Parity operator ($\Gamma_4$) is the Dirac mass term $M = (\varepsilon(\mathbf{k}) - \varepsilon(\mathbf{k}+\mathbf{Q}))/2$. Therefore, the topological invariant can be evaluated by the parity analysis [Eq. (16)] by tracking the band inversion induced by the main and folded bands at the TR invariant points.

The physical interpretation of the SODW induced TI is that the electron-electron interaction induces a chirality inversion in the real-space between different sublattices. In 2D, the emergent QSH effect is therefore spatially modulated as demonstrated in Fig. 16, which we call quantum spin-Hall density wave (QSHDW) insulator. In typical topological classes, topological invariants arise from the non-trivial geometry of the band topology of non-interacting fermions. Electron-electron interaction does not directly drive a topological phase transition, except in few cases such as topological Mott, or Kondo or Anderson insulators, in which, however, no Landau-type order parameter develops. The proposed QSHDW phase is a new kind of Landau order parameter which is associated with topological invariant. The realization of this order state requires SOC, strong coulomb interaction, as well as chemical potential tunability to obtain the desired nesting wavevector. Quantum wires of Pb, Bi and other SOC elements are ideal systems to study this problem.[128] Because here due to quasi-1D nature, Fermi surface nesting is enhanced, and here all these three parameters can be tuned both externally and internally. The QSHDW phase can also be explored in non-centrosymmetric heavy-fermion materials which allow SOC split band structure.

### D. Spinless orbital texture inversion induced topological 'chiral orbits'

So far we have discussed the formation of magnetic field free chiral electrons in two methods, namely via the staggered hopping in SSH model[15] or in graphene lattice,[94] and due to SOC. In some of the nodal superconductors, such as nodal *d*-wave copper-oxide superconductors, owing to the particle-hole symmetry, the quasiparticle dispersion around the discrete nodal point acts as Dirac excitations.[85,86] Therefore, in the existing classes of Dirac materials, 'massless' Dirac fermions only appears in certain conditions. For example, Dirac cones only form in atomically thin layer of graphene, or on the surface states inside the bulk gap in TI. Therefore, the relevant materials choices are restricted to heavy elements which, by nature, have lower band velocity and higher correlation strength.

Recently, the present author have proposed a new theory for a distinct type of Dirac materials, called 'Weyl/Dirac orbital semimetals' and topological orbital insulator, which has Weyl/Dirac cone arising from the orbital texture inversion at discrete momenta between two orbitals with different symmetries.[16,17] The general idea of this design principle lies in assembling different atoms with distinct conduction electrons in such a way that the inter-orbital electron hopping or tunneling term naturally obtains an odd function of energy-momentum dispersion. Finally, the intra-orbital terms also conspire in such a way that the corresponding low-energy Hamiltonian can be reduced to an effective **k.p** - type Dirac Hamiltonian. The resulting Dirac/ Weyl cones at the orbital degenerate points are protected by lattice/ translational symmetry.

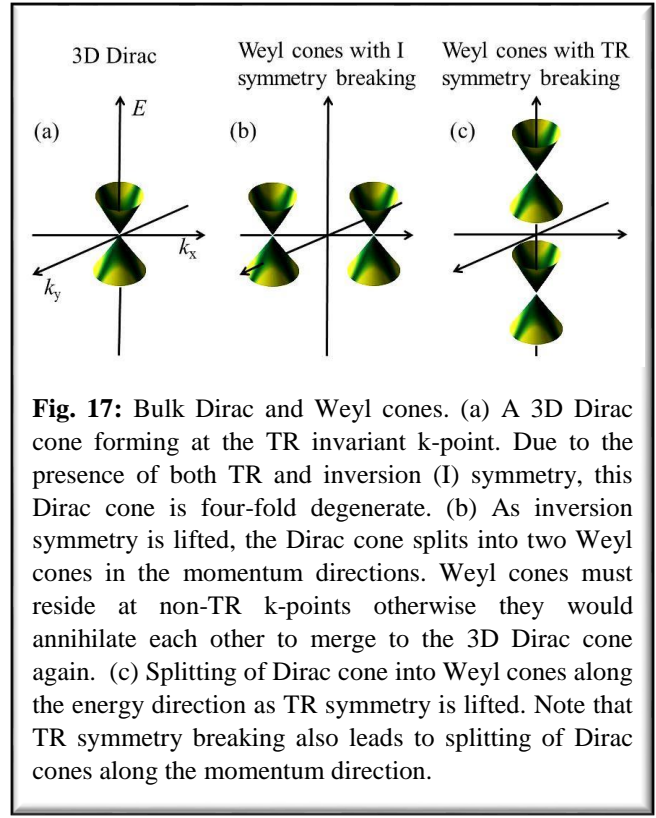

**Fig. 17:** Bulk Dirac and Weyl cones. (a) A 3D Dirac cone forming at the TR invariant k-point. Due to the presence of both TR and inversion (I) symmetry, this Dirac cone is four-fold degenerate. (b) As inversion symmetry is lifted, the Dirac cone splits into two Weyl cones in the momentum directions. Weyl cones must reside at non-TR k-points otherwise they would annihilate each other to merge to the 3D Dirac cone again. (c) Splitting of Dirac cone into Weyl cones along the energy direction as TR symmetry is lifted. Note that TR symmetry breaking also leads to splitting of Dirac cones along the momentum direction.

We discuss here a specific example for engineering Weyl cones. Weyl cones are split Dirac cones arising from the Dirac Hamiltonian when either inversion or TR symmetry is lifted, see Fig. 17. We take a layer-by-layer setup which includes even and odd parity orbitals in alternating layers - dubbed orbital selective superlattice. Such structure is odd under mirror symmetry along the superlattice growth axis. As shown in Fig. 18, we consider '*s*' and '$p_z$' orbitals, placed along the z-direction in different layers, such that inter-orbital or inter-layer hopping has same amplitude (unlike in the SSH model[15] in which the amplitude itself is different), but acquires different sign. The net tunneling matrix-element then becomes purely imaginary as $\varepsilon^{sp}(\mathbf{k}) = -2it_z^{sp}\sin(k_z c)$, where $t_z^{sp}$ is the hopping amplitude and *c* is the interlayer distance. Such complex hopping term can also arise in various other orbital combinations, such as a combination of bonding and antibonding states, or from the mixture of two even orbitals (such as *s* and $d_{xy}$), or two odd orbitals (such as $p_x$ and $p_y$) orbitals when they are placed diagonally, as discussed in Ref. [17]. Let us fix $d_3(k) = \varepsilon^{sp}(k_z)$, hence we are left with obtaining imaginary in-plane hoppings corresponding to $d_1$ and $d_2$. Here we can seek the same principle as orbital selective in-plane lattice sites, or a hexagonal lattice, or $p_x + p_y$- wave pairing or SOC, all yielding essentially similar Hamiltonians. We take a Rashba-type SOC which sets $d_1 = \alpha_R\sin(k_xa)$, and $d_2 = -i\alpha_R\sin(k_ya)$. This Rashba-SOC is the lattice generalization of the same Rashba-SOC discussed in Sec. IIIC. Now the Hamiltonian can be expressed in the usual Γ-matrix form in the orbital spinor $\Psi_k = (\psi_{k\uparrow}^s, \psi_{k\downarrow}^s, \psi_{k\uparrow}^p, \psi_{k\downarrow}^p)$ as

$$H(\mathbf{k}) = \varepsilon^+(\mathbf{k})\mathbb{I}_{4\times4} + \varepsilon^-(\mathbf{k})\Gamma_4 + \mathbf{d}(\mathbf{k})\cdot\mathbf{\Gamma}, \quad (30)$$

where $\varepsilon^\pm(\mathbf{k}) = (\varepsilon^s(\mathbf{k})\pm\varepsilon^p(\mathbf{k}))/2$, with $\varepsilon^{s/p}(\mathbf{k})$ being the intra-orbital dispersions. The energy spectrum is $E^\pm(\mathbf{k}) = \varepsilon^+(\mathbf{k}) \pm \sqrt{\left(\varepsilon^-(\mathbf{k})\right)^2 + |\mathbf{d}(\mathbf{k})|^2}$. Therefore, the contour of $\varepsilon^-(\mathbf{k}) = 0$ gives a gapless nodal ring, among which all the k-points become gapped by finite values of **d**-vectors except those discrete points at which all its components vanish simultaneously. At these discrete *k*-points, Dirac/Weyl cones arise. $\Gamma_4$ is the parity operator, and is even under TR, while $\Gamma_{1,2,3}$ are odd under TR. Since $\varepsilon^-(\mathbf{k})$ is even function of **k**, while $d_{1,2,3}$ are odd functions in momentum, the Hamiltonian is invariant under both inversion and TR symmetries. In such case, 3D Dirac cones can appear only at the TR invariant high-symmetric k-points, and are four-fold degenerate. As the inversion symmetry is lifted, they split in the momentum space, while with TR symmetry breaking, the splitting can occur in either momentum or energy direction. The corresponding two-fold degenerate nodal cones are called the Weyl cones, each enclosing integer, but opposite Chern numbers. Therefore, Weyl cones always appear in pairs, each representing the center of counter helical 'chiral orbits', but remain protected by translational symmetry.[135] As a negative band gap is opened at the 3D Dirac cone, the system becomes a $Z_2$ TI. Similarly, negative band gap at the Weyl nodes can also give rise to topological crystalline insulators if mirror symmetry is present. Such orbital texture inversion induced TIs is refereed as topological orbital insulators. One can envision to engineer topological orbital insulators in a similar fashion discussed in Sec. IIIB, with layer-by-layer approach, in which stacking adjacent layers must host Weyl/Dirac cones with opposite chirality at the same momentum.

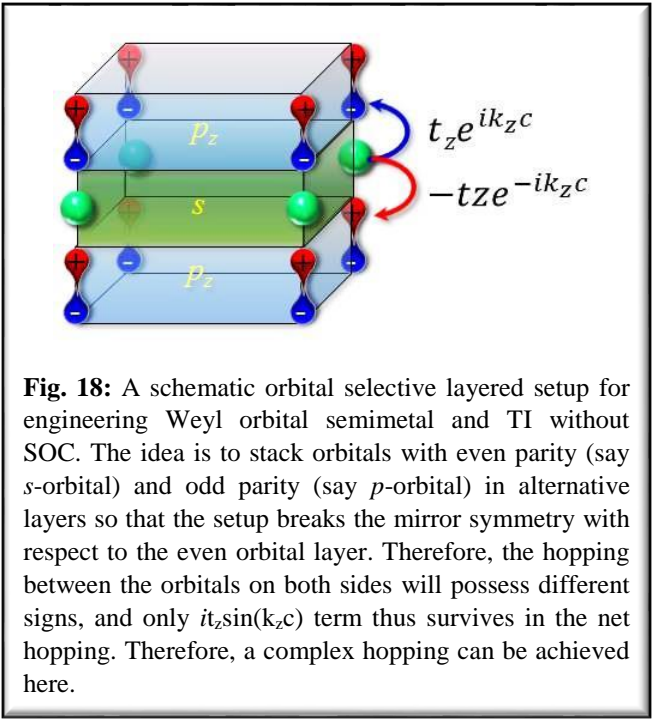

**Fig. 18:** A schematic orbital selective layered setup for engineering Weyl orbital semimetal and TI without SOC. The idea is to stack orbitals with even parity (say *s*-orbital) and odd parity (say *p*-orbital) in alternative layers so that the setup breaks the mirror symmetry with respect to the even orbital layer. Therefore, the hopping between the orbitals on both sides will possess different signs, and only $it_z\sin(k_zc)$ term thus survives in the net hopping. Therefore, a complex hopping can be achieved here.

Recently, we have predicted that the ferromagnetic $V_3S_4$ is an intrinsic Weyl orbital semimetal.[17] The theory of Weyl orbital semimetals does not depend on the uncommon conditions such as sublattice symmetry in atomically thin graphene, or high value of SOC in TIs. Engineering Weyl orbital materials will expand the territory of the Dirac materials beyond the typical heavy elements' Dirac systems or graphene to even lower atomic number system and thereby enhance the value of Fermi velocity. The nature of impurity scattering protection in this case is characteristically different. In the present family, an electron can only scatter from one orbital state to another when the impurity vertex contains a corresponding anisotropic orbital-exchange matrix- element or if the electron dynamically passes through the momentum and energy of the Dirac cone. Another advantage of the Weyl orbital semimetal is that here the Dirac cone is even immune to TR symmetry breaking, and a bulk gap can be engineered by the lattice distortion. Therefore, the generation, transport and detection of orbitally protected electric current may lead to new opportunities for orbitronics. Chiral orbital current in the Weyl semimetals can be detected by Kerr effect.

## IV. Conclusion and outlook

This article presented a thorough understanding of the topological invariant in the absence and presence of TR symmetry, as well as other symmetries such as inversion symmetry, mirror symmetry, or particle-hole symmetry. The key strategy, that we followed here, is that the historical development of the topological invariant starting from the quantization of flux through a non-trivial Gaussian curvature to the IQH to QSH to $Z_2$ TI provides a more transparent and step-by-step development of this field. A unique feature of this field is that it blends concepts from various fields including mathematics, condensed matter physics, chemistry, and particle physics. A detailed discussion of the materials chemistry of the TI systems is left out in this article and can be found elsewhere.[63] We can highlight an interesting difference between the trivial insulator and TI from a chemist's perspective. In a band insulator, the valence and conduction bands remain fully filled and empty, respectively. Therefore, such insulators are sought in elements or compounds whose outer most orbital contains even number of electrons. TI is also a 'band insulator' with a twist. Here the valence and conduction bands are also expected to be completely filled and empty, respectively, but at the same time, these bands are inverted at the TR invariant k-points, suggesting otherwise. Therefore, one can neither seek for elements with completely full outermost orbital, nor partially filled orbital (since partially filled state should give a metallic state). For the same reason, the material should neither be too covalent, nor too ionic.

Therefore intermetallics are obvious elements to consider for non-interacting TI (interaction can change this simple explanation). Indeed, most of the TIs are made of intermetallics. Here we can consider at least two partially occupied elements which participate in orbital-overlap and/or coupled via SOC. Therefore, a band can remained fully occupied by accommodating two partially filled orbitals. If the exchange of the orbital character occurs at odd number of TR invariant points, it gives rise to 'strong' TI, otherwise even number of orbital weight switching can give rise to a either a 'weak' TI or a trivial insulator.

From high-energy physicists' perspective, TI field offers a plenty of new opportunities to predict new excitation as well as to realize some of the uncharted 'particles' predicted there. Weyl fermions[136–139] and Majorana fermions[140–142] which remained elusive for decades have only been realized recently in TI platforms. Axion,[24,104] and anyons[143,144] are two widely searched excitations which are predicted to be present in TIs. Recently, it is shown that supersymmetry (SUSY) can be found in TI and superconductor heterostructures.[145,146]

Cold atom physicists also find it interesting to contribute to the TI fields in various ways. Haldane, in his original paper,[3] commented that "the particular model presented here is unlikely to be directly physically realizable". Cold atom researchers have made it possible to create the Haldane model with optically generated honeycomb lattice.[96] They have engineered TR and inversion symmetry breakings by carefully enhancing next-nearest-neighbor hopping and providing staggered onsite energies to different sublattices, respectively, as proposed by Haldane. More recently, protected edge state is generated in various optical lattice structure by employing synthetic gauge field.[147,148] The apparent setback to realize the $Z_2$ TI due to the lack of 2D SOC can be overcome with our engineered structure (as discussed in Sec. IIIA),[79] and the realization of TR invariant TI in optical lattice field is a matter of time. Due to tremendous controls over structural and quantum properties in these setups, an unprecedented tunability of TI properties for further exploration can be possible here.

Despite the predictions and discoveries of several TI materials classes, materials flexibility still remains a grand challenge.[63] $Bi_2Se_3$ family is widely used for many experiments on TI since both single crystal, and thin films of this system can be easily grown. HgTe/CdTe[33] and InAs/GaSb[72] are the only two systems experimentally demonstrated as QSH insulators, which are however not being used in other experiments. Magnetic doped thin film of $Bi_2Se_3$ is the only system synthesized so far to be QAH insulator.[107,108] $Pb_{1-x}Sn_xSe/Te$[52–54] and SnS[55] are the only two systems known to be topological crystalline insulator. $SmB_6$ is predicted and subsequently realized to be topological Kondo insulator,[56,149,150] although evidence against this conclusion is also present.[151] $Cd_3As_2$,[152,153] and $Na_3Bi$[154,155] are synthesized to be 3D TI, while TaAs,[136–138] and NbAs[139] family is discerned recently to be Weyl semimetal. Many other families, such as topological Mott insulator, topological Anderson insulator, and topological axion insulators are yet to be discovered. Moreover, the inevitable presence of the bulk conductivity in most of the 3D TI samples poses a serious nuisance to experimentalists. On the other hand, engineering TI can be rather simpler. It also offers tremendous versatility in terms of materials growth, and obtaining quantum and topological properties. Therefore, the successful preparation of 'home-made' TI will caters to many physics, chemistry and engineering fields seeking suitable materials with higher tunability and materials flexibility.

## Acknowledgments

I have many people to thank for their generous collaborations, discussions, and lessons on various aspects of topological states of matter and other related concepts which led to the completion of this article. There is a possibility that I fail to acknowledge everybody here to the rush in finishing this article. The people I can immediately recall to convey my gratitude are: Arun Bansil, Alexander Balatsky, Hsin Lin, Le-Quy Duong, F.-C. Chuang, M. Zahid Hasan, Diptiman Sen, Anindya Das, Aveek Bid, Tirupathaiah Setti, Vidya Madhavan, Robert Markiewicz, Matthias Graf, Jian-Xin Zhou, Christoph Tegenkamp, J. Hugo Dil. Kapildeb Dolui, Gaurav Gupta, Sujay Ray, Ananya Ghatak, Sayonee Ray, Kallol Sen, Soumi Ghosh, Kausik Ghosh.